\documentclass[journal]{IEEEtran}

\usepackage{amssymb, latexsym}
\usepackage{amsmath} 
\usepackage{amsthm}
\usepackage{bm}
\usepackage{graphicx}
\usepackage{epstopdf}  
\usepackage[mathscr]{eucal}
\usepackage{cite}
\usepackage{multirow}
\usepackage{subfig}

\graphicspath{{./Figures/}{./}}

\ifCLASSINFOpdf
\else
\fi

\begin{document}
\setlength{\abovedisplayskip}{6pt}
\setlength{\belowdisplayskip}{6pt}

%
% paper title
% Titles are generally capitalized except for words such as a, an, and, as,
% at, but, by, for, in, nor, of, on, or, the, to and up, which are usually
% not capitalized unless they are the first or last word of the title.
% Linebreaks \\ can be used within to get better formatting as desired.
% Do not put math or special symbols in the title.
\title{Travel time tomography with adaptive dictionaries}
%
%
% author names and IEEE memberships
% note positions of commas and nonbreaking spaces ( ~ ) LaTeX will not break
% a structure at a ~ so this keeps an author's name from being broken across
% two lines.
% use \thanks{} to gain access to the first footnote area
% a separate \thanks must be used for each paragraph as LaTeX2e's \thanks
% was not built to handle multiple paragraphs
%

\author{Michael~Bianco,~\IEEEmembership{Student Member,~IEEE,}
        Peter~Gerstoft,~\IEEEmembership{Member,~IEEE,}
\thanks{M.~Bianco and P.~Gerstoft are with the Scripps Institution of Oceanography, University of California San Diego, La Jolla, CA 92093-0238, USA, http://noiselab.ucsd.edu (e-mail: mbianco@ucsd.edu; gerstoft@ucsd.edu)}
\thanks{Supported by the Office of Naval Research, Grant No. N00014-18-1-2118.}% <-this % stops a space
%\thanks{Manuscript received \#\#\#\#\#; revised \today.}}
}

% note the % following the last \IEEEmembership and also \thanks - 
% these prevent an unwanted space from occurring between the last author name
% and the end of the author line. i.e., if you had this:
% 
% \author{....lastname \thanks{...} \thanks{...} }
%                     ^------------^------------^----Do not want these spaces!
%
% a space would be appended to the last name and could cause every name on that
% line to be shifted left slightly. This is one of those "LaTeX things". For
% instance, "\textbf{A} \textbf{B}" will typeset as "A B" not "AB". To get
% "AB" then you have to do: "\textbf{A}\textbf{B}"
% \thanks is no different in this regard, so shield the last } of each \thanks
% that ends a line with a % and do not let a space in before the next \thanks.
% Spaces after \IEEEmembership other than the last one are OK (and needed) as
% you are supposed to have spaces between the names. For what it is worth,
% this is a minor point as most people would not even notice if the said evil
% space somehow managed to creep in.

% The paper headers
%\markboth{Revised \today}%
\markboth{}%
{Shell \MakeLowercase{\textit{et al.}}: Bare Demo of IEEEtran.cls for IEEE Journals}
% The only time the second header will appear is for the odd numbered pages
% after the title page when using the twoside option.
% 
% *** Note that you probably will NOT want to include the author's ***
% *** name in the headers of peer review papers.                   ***
% You can use \ifCLASSOPTIONpeerreview for conditional compilation here if
% you desire.

% If you want to put a publisher's ID mark on the page you can do it like
% this:
%\IEEEpubid{0000--0000/00\$00.00~\copyright~2015 IEEE}
% Remember, if you use this you must call \IEEEpubidadjcol in the second
% column for its text to clear the IEEEpubid mark.

% use for special paper notices
%\IEEEspecialpapernotice{(Invited Paper)}

% make the title area
\maketitle
%REVISED: \today

% As a general rule, do not put math, special symbols or citations
% in the abstract or keywords.
\begin{abstract}
We develop a 2D travel time tomography method which regularizes the inversion by modeling groups of slowness pixels from discrete slowness maps, called patches, as sparse linear combinations of atoms from a dictionary. We propose to use dictionary learning during the inversion to adapt dictionaries to specific slowness maps. This patch regularization, called the local model, is integrated into the overall slowness map, called the global model. The local model considers small-scale variations using a sparsity constraint and the global model considers larger-scale features constrained using $\ell_2$ regularization. This strategy in a locally-sparse travel time tomography (LST) approach enables simultaneous modeling of smooth and discontinuous slowness features. This is in contrast to conventional tomography methods, which constrain models to be exclusively smooth or discontinuous. We develop a \textit{maximum a posteriori} formulation for LST and exploit the sparsity of slowness patches using dictionary learning. The LST approach compares favorably with smoothness and total variation regularization methods on densely, but irregularly sampled synthetic slowness maps.
\end{abstract}

% Note that keywords are not normally used for peerreview papers.

% For peer review papers, you can put extra information on the cover
% page as needed:
% \ifCLASSOPTIONpeerreview
% \begin{center} \bfseries EDICS Category: 3-BBND \end{center}
% \fi
%
% For peerreview papers, this IEEEtran command inserts a page break and
% creates the second title. It will be ignored for other modes.
\IEEEpeerreviewmaketitle

\section{Introduction}
\label{sec:intro}
Travel time tomography methods estimate Earth slowness structure, which contains smooth and discontinuous features at multiple spatial scales, from travel times of seismic waves between recording stations. The estimation of slowness (inverse of speed) models from travel times is often formulated as a discrete linear inverse problem, where the perturbations in travel time relative to a reference are used to infer the unknown structure\cite{rawlinson2010,aster2013}. Such problems are ill-posed, with irregular ray coverage of environments, and require regularization to obtain physically plausible solutions.

We propose a 2D travel time tomography method which regularizes the inversion by assuming small groups of slowness pixels from a discrete slowness map, called patches, are well approximated by sparse linear combinations of atoms from a dictionary. In this sparse model \cite{elad2010,mairal2014}, the atoms represent elemental slowness patches and can be generic dictionaries, e.g. wavelets, or adapted to specific data by dictionary learning\cite{aharon2006,schnass2015}. This patch regularization, called the local model, is integrated into the overall slowness map, called the global model. Whereas the local model considers small-scale variations using a sparsity constraint, the global model considers larger-scale features which are constrained using $\ell_2$ regularization. 

This local-global modeling strategy with dictionary learning has been successful in image denoising \cite{elad2006,elad2010,mairal2009} and inpainting \cite{mairal2012}, where natural image content is recovered from noisy or incomplete data. We use this strategy to recover true slowness fields from travel time tomography by simultaneously modeling smooth and discontinuous slowness features. This gives an improvement over conventional methods with global damping and smoothness regularization \cite{aster2013,tarantola1987} and pixel level regularization e.g. total variation (TV) regularization \cite{lin2015,zhang2017} which regularize tomography by encouraging smooth or discontinuous slownesses. Relative to existing tomography methods based on wavelets \cite{chiao2001,hawkins2015} and sparse dictionaries \cite{loris2007,loris2010,charlety2013,fang2014}, our formulation of the tomography problem permits the adaptation of the sparse dictionaries to travel time data and ray sampling by dictionary learning techniques, for which many methods exist \cite{olshausen1997,aharon2006,mairal2009,elad2010,schnass2015}. Sparse reconstruction performance is often improved using adaptive dictionaries, which represent well specific data, over generic dictionaries which achieve acceptable performance for many tasks\cite{mairal2014}.

Sparse modeling assumes signals can be reconstructed to acceptable accuracy using few or \textit{sparse} vectors, called \textit{atoms}, from a potentially large set or \textit{dictionary} of atoms. The parsimony of sparse representations \cite{mairal2014} often provides better regularization than, for example, traditional $\ell_2$ model damping\cite{aster2013}. Early sparse approaches were developed in seismic deconvolution \cite{taylor1979,chapman1983}. This philosophy has since become ubiquitous in signal processing for image and video denoising \cite{elad2006,elad2010,mairal2009} and inpainting \cite{mairal2012}, and medical imaging \cite{lustig2007,ravishankar2011}, to name a few examples. Recent works in acoustics and seismics have utilized sparse modeling, e.g.~beamforming \cite{xenaki2014,gerstoft2016}, matched field processing \cite{gemba2017}, estimation of ocean acoustic properties \cite{taroudakis2015,wang2016,bianco2016,bianco2017a,bianco2017b,choo2018}. Dictionary learning has been used to denoise seismic \cite{beckouche2014,chen2017} and ocean acoustic \cite{taroudakis2015} recordings, to regularize full waveform inversion \cite{zhu2017,li2018}, and to regularize ocean sound speed profile inversion \cite{wang2016,bianco2016}.

Inspired by image denoising \cite{elad2006}, we develop a sparse and adaptive 2D travel time tomography method, which we refer to as locally-sparse travel time tomography (LST). Whereas in \cite{elad2006}, the image pixel values are directly observed, in LST the pixel values are inferred from measurements \cite{ravishankar2011}. This necessitates an extra term to fit slowness pixels to travel time observations. We develop a \textit{maximum a posteriori} (MAP) formulation for LST and use the iterative thresholding and signed K-means (ITKM)\cite{schnass2015} dictionary learning algorithm to design adaptive dictionaries. This improves slowness models over generic dictionaries. We demonstrate the performance of LST for 2D surface wave tomography with synthetic slowness maps and travel time data. The LST results compare favorably with two competing methods: a smoothing and damping approach\cite{rodgers2000} we here deem conventional tomography, and TV regularization\cite{lin2015}.

\section{LST model formulation}
\label{sec:overview}
In developing the LST method, we consider the case of 2D travel time tomography, where slowness of the medium varies only in two dimensions. In the case of seismic tomography, surface wave tomography is one case where this assumption is valid\cite{barmin2001}. The sensing configuration for such a scenario in Fig.~\ref{fig:patchMap}(a). We discretize a 2D slowness map as a $W_1\times W_2$ pixel image, where the pixels have constant slowness. An array of sensors in the 2D map register waves propagating across the array. From these observations, wave travel times between the sensors, $\mathbf{t'}\in\mathbb{R}^M$, are obtained. We assume $\mathbf{t'}$ given and disregard refraction of the waves, yielding a `straight-ray' formulation of the problem. Such rays are illustrated in Fig.~\ref{fig:patchMap}(a). The tomography problem is to estimate the slowness pixels (see Fig.~\ref{fig:patchMap}(a)) from $\mathbf{t'}$. 

In the following, we develop separately two slowness models, deemed the \textit{global} and \textit{local} models, which will be related in Section~\ref{sec:propMAP}, and briefly discuss dictionaries for sparse modeling. The global model considers the larger scale or global features and relates travel times to slowness. The local model considers smaller scale or more localized features with sparse modeling.

\subsection{Global model and travel times}
In the global model, slowness pixels (see Fig.~\ref{fig:patchMap}(a)) are represented by the vector $\mathbf{s'=s}_\mathrm{g}+\mathbf{s}_0\in\mathbb{R}^{N}$, where $\mathbf{s}_0$ is reference slownesses and $\mathbf{s}_\mathrm{g}$ is perturbations from the reference, here referred to as the {\it global slowness}, with $N=W_1W_2$. Similarly, the travel times of the $M$ rays are given as $\mathbf{t'=t+t}_0$, where $\mathbf{t}$ is the travel time perturbation and $\mathbf{t}_0$ is the reference travel time. The tomography matrix $\mathbf{A}\in\mathbb{R}^{M\times N}$ gives the discrete path lengths of $M$ straight-rays through $N$ pixels (see Fig.\ \ref{fig:patchMap}(a)). Thus  $\mathbf{t}$ and $\mathbf{s}_\mathrm{g}$ are related by the linear measurement model
%
%\begin{linenomath*}
\begin{equation}
\mathbf{t}=\mathbf{As}_\mathrm{g}+\epsilon,
\label{eq:linearTraveltime}
\end{equation}
%\end{linenomath*}
%
where $\mathbf{\epsilon}\in\mathbb{R}^M$ is Gaussian noise $\mathcal{N}(\mathbf{0},\sigma_\epsilon^2\bf{I})$, with mean $\mathbf{0}$ and covariance $\sigma_\epsilon^2\bf{I}$. We estimate the perturbations, with $\mathbf{s}_0$ and $\mathbf{t}_0=\mathbf{A}\mathbf{s}_0$ known. We call (\ref{eq:linearTraveltime}) the \textit{global model}, as it captures the large-scale features that span the discrete map and generates $\mathbf{t}$. 

\subsection{Local model and sparsity}
\begin{figure}[t]
\centering
\includegraphics[width=6cm]{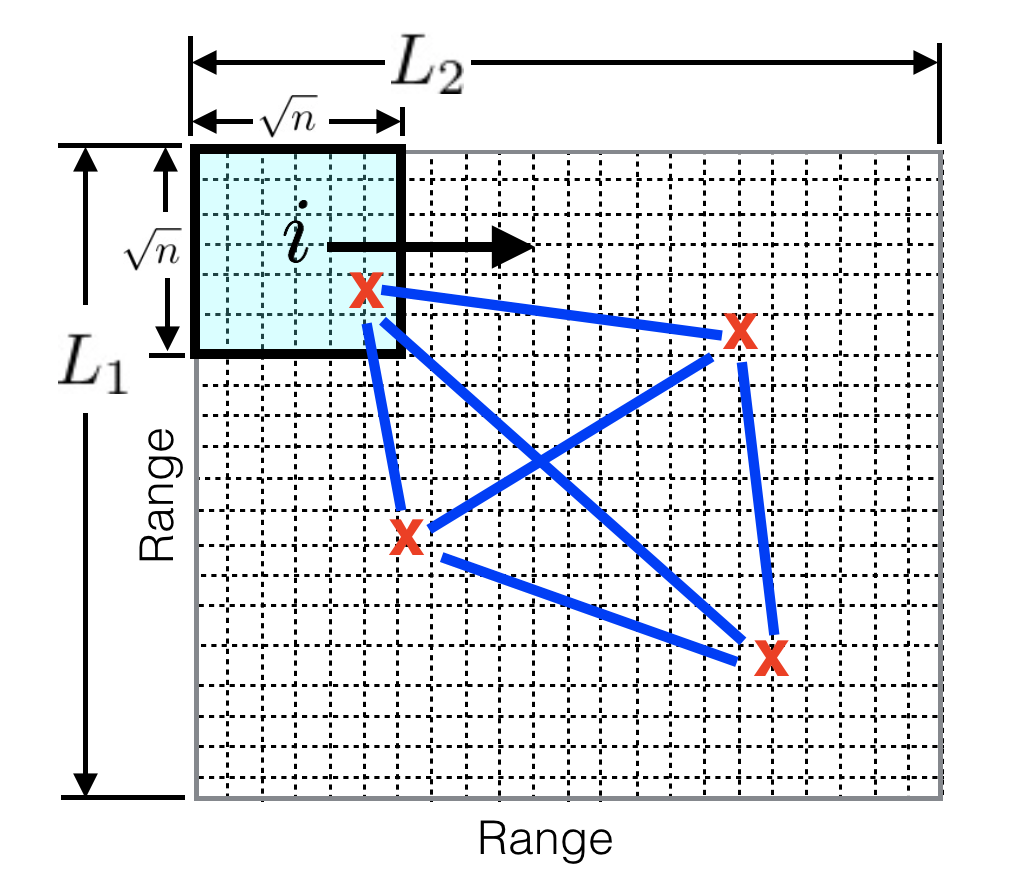} \\
\caption{2D slowness image corresponds to slowness map divided into pixels (dashed boxes). The square image patch $i$ contains $n$ pixels. $W_1$ and $W_2$ are the vertical and horizontal dimensions of the image (in pixels), which give $I$ unique patches. Sensors are shown as red x's and the ray paths between the sensors are shown as blue lines.}
\label{fig:patchMap}
\end{figure}

In the local model, slowness pixels (see Fig.~\ref{fig:patchMap}(a)) are represented by the vector $\mathbf{s'=s}_\mathrm{s}+\mathbf{s}_0\in\mathbb{R}^{N}$, where $\mathbf{s}_\mathrm{s}$ is perturbations from the reference, here referred to as the {\it sparse slowness}. We assume that {\it patches}, or $\sqrt{n}\times\hspace{-0.4ex}\sqrt{n}$ groups of pixels from $\mathbf{s}_\mathrm{s}$ (see Fig.\ \ref{fig:patchMap}(a)) are well approximated by a sparse linear combination of atoms from a dictionary $\mathbf{D}\in\mathbb{R}^{n\times Q}$ of $Q$ atoms. The patches are selected from $\mathbf{s}_\mathrm{s}$ by the binary matrix $\mathbf{R}_i\in\{0,1\}^{n\times N}$. Hence the slownesses in patch $i$ are $\mathbf{R}_i\mathbf{s}_\mathrm{s}$. The sparse model is formulated as 
%\begin{linenomath*}
\begin{equation}
\mathbf{R}_i\mathbf{s}_\mathrm{s}\approx\mathbf{D}\mathbf{x}_i~\text{and}\ |\mathbf{x}_{i}\ne0|=T~\forall~i
\label{eq:patchForm1}
\end{equation}
%\end{linenomath*}
where $|\cdot|$ is cardinality, and $\mathbf{x}_i\in\mathbb{R}^{n}$ is the sparse coefficients, and $T\ll n$ is the number of non-zero coefficients.  The quantity $\mathbf{D}\mathbf{x}_i$ is referred to as the {\it patch slowness}. We call (\ref{eq:patchForm1}) the {\it local model}, as it captures the smaller scale, localized features contained by patches. 

Each slowness patch $\mathbf{R}_i\mathbf{s}_\mathrm{s}$ is indexed by the row $w_1$ and column $w_2$ of its top-left pixel in the 2D image as $(w_{1,i},w_{2,i})$. We consider all overlapping patches, with $w_{1,i}$ and $w_{2,i}$ differing from their neighbor by $\pm1$ (\textit{stride} of one). Further, the patches wrap-around the edges of the image \cite{aharon2008}. Thus, for a $N=W_1\times W_2$ pixel image, the number of patches $I=N$, and the number of patches per pixel $b=n$ is the same for all pixels.

The atoms in $\mathbf{D}$ are considered ``elemental patches", where only a small number of atoms are necessary to adequately approximate $\mathbf{R}_i\mathbf{s}_\mathrm{s}$. Atoms can be generic functions, e.g. wavelets or the discrete cosine transform (DCT), or learned from the data (see Sec.\ \ref{sec:st3dlearn}). An example of DCT atoms are shown in Fig. \ref{fig:dct}. Adaptive dictionaries, which are designed from specific instances of data using dictionary learning algorithms, often achieve greater reconstruction accuracy over generic dictionaries. Examples of dictionaries learned from synthetic travel time data (from slowness maps in Fig. \ref{fig:mapsSampling}) are shown in Fig. \ref{fig:learnedDicts}. Relative to generic dictionaries, learned dictionaries can represent the smooth and discontinuous seismic features we are likely to encounter in real inversion scenarios. 
%%%%%%%%%%%%%%%%
%%%%%%%%%

\section{LST MAP objective and evaluation}
\label{sec:propMAP}
We now derive a MAP objective for the LST variables and an algorithm for its evaluation, with the ultimate goal of estimating the sparse slowness $\mathbf{s}_\mathrm{s}$ (from (\ref{eq:patchForm1})). Assuming the travel times $\mathbf{t}$, tomography matrix $\mathbf{A}$, and dictionary $\mathbf{D}$ known, the solution to the objective gives MAP estimates of the global slowness $\mathbf{s}_\mathrm{g}$ (from (\ref{eq:linearTraveltime})), $\mathbf{s}_\mathrm{s}$, and the coefficients $\mathbf{X=[x}_1,...,\mathbf{x}_I]\in\mathbb{R}^{Q\times I}$ describing all patches I (from (\ref{eq:patchForm1})). Since we use a non-Bayesian dictionary learning algorithm (ITKM \cite{schnass2015}), dictionary learning is added after the MAP derivation in Sec.\ \ref{sec:st3dlearn}.

\subsection{Derivation of MAP objective}
Starting Bayes' rule, we combine the global (\ref{eq:linearTraveltime}) and local (\ref{eq:patchForm1}) models, formulating the posterior density of the LST variables as
%\vspace{-1ex}
%\begin{linenomath*}
\begin{align}
p\big(\mathbf{s}_\mathrm{g},\mathbf{s}_\mathrm{s},\mathbf{X \big|t}\big) \propto p\big(\mathbf{t}\big|\mathbf{s}_\mathrm{g},\mathbf{s}_\mathrm{s},\mathbf{X}\big)p\big(\mathbf{s}_\mathrm{g},\mathbf{s}_\mathrm{s},\mathbf{X}\big).
\label{eq:bayesRule3_0}
\end{align}
%\end{linenomath*}
From (\ref{eq:patchForm1}), $\mathbf{s}_\mathrm{s}$ is conditioned only on $\mathbf{X}$. We further assume the patch coefficients $\mathbf{X}$ independent. Hence, using the chain rule we obtain from (\ref{eq:bayesRule3_0})
%\begin{linenomath*}
%\vspace{-0.18ex}
\begin{gather}
\begin{aligned}
p\big(\mathbf{s}_\mathrm{g},\mathbf{s}_\mathrm{s},\mathbf{X \big|t}\big) & \propto p\big(\mathbf{t}\big|\mathbf{s}_\mathrm{g},\mathbf{s}_\mathrm{s},\mathbf{X}\big)p\big(\mathbf{s}_\mathrm{g}\big|\mathbf{s}_\mathrm{s},\mathbf{X}\big)p\big(\mathbf{s}_\mathrm{s},\mathbf{X}\big) \\
&\propto p\big(\mathbf{t}\big|\mathbf{s}_\mathrm{g},\mathbf{s}_\mathrm{s},\mathbf{X}\big)p\big(\mathbf{s}_\mathrm{g}\big|\mathbf{s}_\mathrm{s},\mathbf{X}\big)p\big(\mathbf{s}_\mathrm{s}\big|\mathbf{X}\big)p\big(\mathbf{X}\big),
\end{aligned} \raisetag{1.8\baselineskip}
\label{eq:bayesRule3}
\end{gather}
%\end{linenomath*}
From (\ref{eq:linearTraveltime}), $\mathbf{t}$ is conditioned only on $\mathbf{s}_\mathrm{g}$ and we assume $\mathbf{s}_\mathrm{g}$ is conditioned only on $\mathbf{s}_\mathrm{s}$. Hence, we obtain from (\ref{eq:bayesRule3})
%\begin{linenomath*}
\begin{equation}
p\big(\mathbf{s}_\mathrm{g},\mathbf{s}_\mathrm{s},\mathbf{X \big|t}\big) \propto p\big(\mathbf{t}\big|\mathbf{s}_\mathrm{g}\big)p\big(\mathbf{s}_\mathrm{g}\big|\mathbf{s}_\mathrm{s})p\big(\mathbf{s}_\mathrm{s}\big|\mathbf{X}\big)p\big(\mathbf{X}\big),
\label{eq:bayesRule4}
\end{equation}
%\end{linenomath*}

We approximate $p\big(\mathbf{t}\big|\mathbf{s}_\mathrm{g}\big)$, $p\big(\mathbf{s}_\mathrm{g}\big|\mathbf{s}_\mathrm{s}\big)$, $p\big(\mathbf{s}_\mathrm{s} \big|\mathbf{X}\big)$ as Gaussian, and all patch slownesses from (\ref{eq:patchForm1}) independent, giving
%\begin{linenomath*}
\begin{equation}
\begin{aligned}
p\big(\mathbf{t \big|\mathbf{s}_\mathrm{g}}\big)&=\mathcal{N}(\mathbf{A}\mathbf{s}_\mathrm{g},\mathbf{\Sigma}_\epsilon), \\
p\big(\mathbf{s}_\mathrm{g} \big|\mathbf{s}_\mathrm{s}\big)&=\mathcal{N}(\mathbf{s}_\mathrm{s},\mathbf{\Sigma}_\mathrm{g}),\\
p\big(\mathbf{s}_\mathrm{s} \big|\mathbf{X}\big)&=\prod_i p\big(\mathbf{R}_i\mathbf{s}_\mathrm{s} \big|\mathbf{x}_i\big)=\prod_i\mathcal{N}\big(\mathbf{Dx}_i,\mathbf{\Sigma}_{p,i}\big),
\label{eq:gauss12}
\end{aligned}
\end{equation}
%\end{linenomath*}
where $\mathbf{\Sigma}_\epsilon\in\mathbb{R}^{K\times K}$ is the covariance of the travel time error, $\mathbf{\Sigma}_\mathrm{g}\in\mathbb{R}^{N\times N}$ is the covariance of $\mathbf{s}_\mathrm{g}$, and $\mathbf{\Sigma}_{p,i}\in\mathbb{R}^{n\times n}$ is the covariance of the patch slownesses. Taking the logarithm of the conditional probabilities from (\ref{eq:gauss12}), we obtain
%\begin{linenomath*}
\begin{equation}
\begin{split}
\ln p\big(\mathbf{t} \big|\mathbf{s}_\mathrm{g}\big)&\propto-\frac{1}{2}(\mathbf{t-A}\mathbf{s}_\mathrm{g})^\mathrm{T}\mathbf{\Sigma}_\epsilon^{-1}(\mathbf{t-A}\mathbf{s}_\mathrm{g}), \\
\ln p\big(\mathbf{s}_\mathrm{g}\big|\mathbf{s}_\mathrm{s}\big)&\propto-\frac{1}{2}(\mathbf{s}_\mathrm{g}-\mathbf{s}_\mathrm{s})^\mathrm{T}\mathbf{\Sigma}_\mathrm{g}^{-1}(\mathbf{s}_\mathrm{g}-\mathbf{s}_\mathrm{s}), \\
\ln p\big(\mathbf{s}_\mathrm{s} \big|\mathbf{X}\big)&\propto-\frac{1}{2}\sum_i(\mathbf{D}\mathbf{x}_{i}-\mathbf{R}_i\mathbf{s}_\mathrm{s})^\mathrm{T}\mathbf{\Sigma}_{p,i}^{-1}(\mathbf{D}\mathbf{x}_{i}-\mathbf{R}_i\mathbf{s}_\mathrm{s}).
\label{eq:lngauss12}
\end{split}\raisetag{3\baselineskip}
\end{equation}
%\end{linenomath*}
%%%%%%%%%%%%%%%%%%%%%%%%
\begin{figure}[t]
\vspace{-2ex}
\hspace{-2ex}\includegraphics[width=9cm]{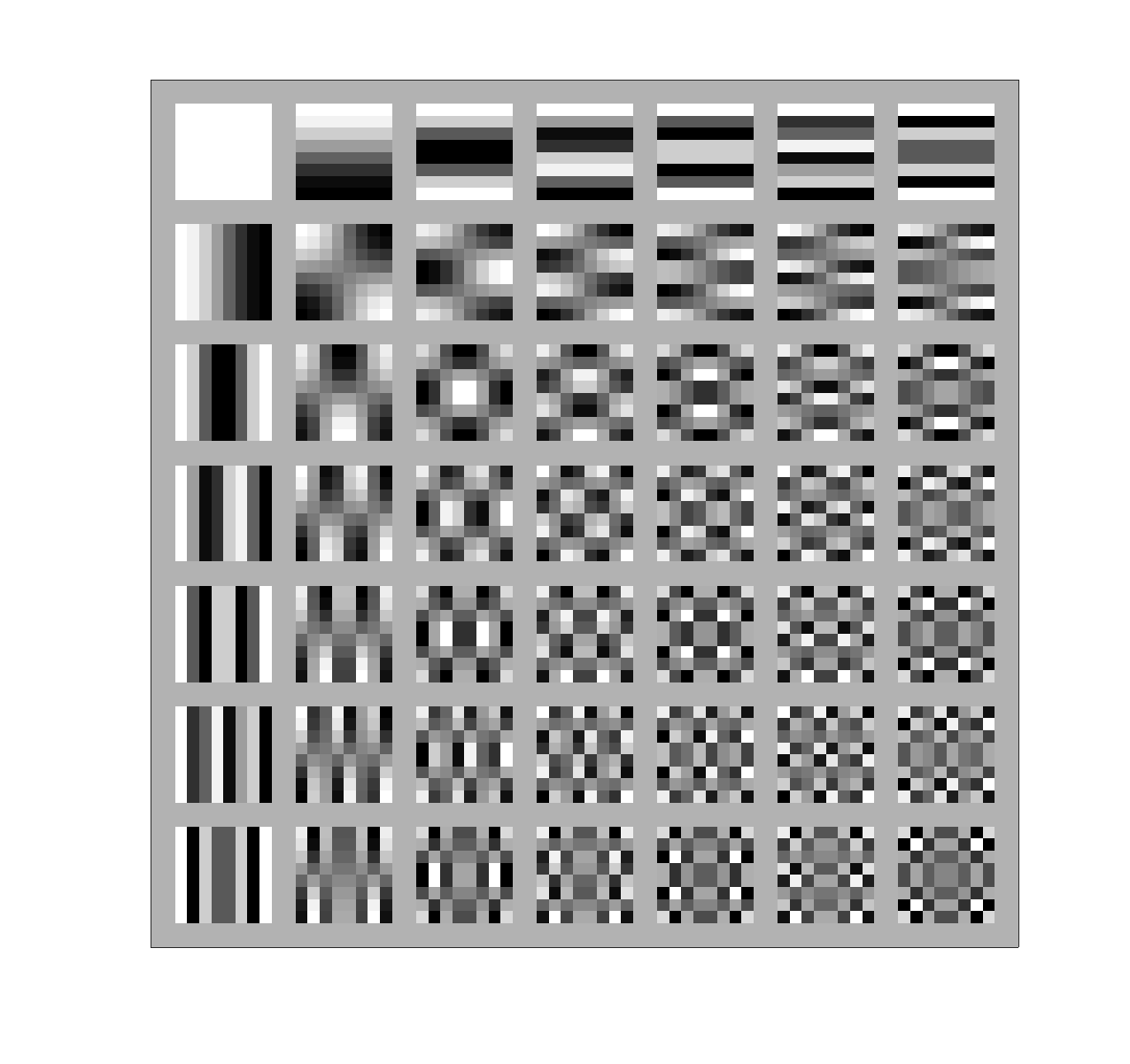}\vspace{-2ex}
\caption{Discrete cosine transform (DCT) dictionary atoms. The $Q=49$ atoms (ordered in a $7\times7$ grid) each  contain $n=8\times8=64$ pixels. Atom values stretched to full grayscale range for display.}
\label{fig:dct}
\end{figure}
%%%%%%%%%%%%%%%%%%%%%%%%

Assuming the coefficients $\mathbf{x}_i$ describing patch $\mathbf{R}_i\mathbf{s}_\mathrm{s}$ are independent,
%\begin{linenomath*}
\begin{equation}
\ln p\big(\mathbf{X}\big)=\sum_i\ln p\big(\mathbf{x}_i\big).
\label{eq:patchPriorLog}
\end{equation}
%\end{linenomath*}

From (\ref{eq:bayesRule4}), and (\ref{eq:lngauss12}), and (\ref{eq:patchPriorLog})
%\begin{linenomath*}
\begin{equation}
\begin{split}
&\ln p\big(\mathbf{s}_\mathrm{g},\mathbf{s}_\mathrm{s},\mathbf{X \big|t}\big) \propto \ \ln \big\{p\big(\mathbf{t}\big|\mathbf{s}_\mathrm{g}\big)p\big(\mathbf{s}_\mathrm{g}\big|\mathbf{s}_\mathrm{s})p\big(\mathbf{s}_\mathrm{s}\big|\mathbf{X})p\big(\mathbf{X}\big)\big\} \\
&\propto -(\mathbf{t-As}_\mathrm{g})^\mathrm{T}\mathbf{\Sigma}_\epsilon^{-1}(\mathbf{t-As}_\mathrm{g})-(\mathbf{s}_\mathrm{g}-\mathbf{s}_s)^\mathrm{T}\mathbf{\Sigma}_\mathrm{g}^{-1}(\mathbf{s}_\mathrm{g}-\mathbf{s}_\mathrm{s}) \\
&-\sum_i\bigg\{(\mathbf{D}\mathbf{x}_{i}-\mathbf{R}_i\mathbf{s}_\mathrm{s})^\mathrm{T}\mathbf{\Sigma}_{p,i}^{-1}(\mathbf{D}\mathbf{x}_{i}-\mathbf{R}_i\mathbf{s}_\mathrm{s})+2\ln p\big(\mathbf{x}_{i}\big)\bigg\}.
\end{split}\raisetag{4\baselineskip}
\label{eq:lnBayes3}
\end{equation}
%\end{linenomath*}
Assuming the coefficients $\mathbf{x}_i$ sparse, we approximate $\ln p\big(\mathbf{x}_{i}\big)$ with the $\ell_0$ pseudo-norm $\|\mathbf{x}_i\|_0$, which counts number of non-zero coefficients \cite{elad2010}. We further assume the number of non-zero coefficients $T$ is the same for every patch. This gives the MAP objective as
%\begin{linenomath*}
\begin{equation}
\begin{split}
&\max\big\{ \ln p\big(\mathbf{s}_\mathrm{g},\mathbf{s}_\mathrm{s},\mathbf{X \big|t}\big)\big\}=\min\big\{ -\ln p\big(\mathbf{s}_\mathrm{g},\mathbf{s}_\mathrm{s},\mathbf{X \big|t}\big)\big\} \\
&\propto\min\bigg\{(\mathbf{t-As}_\mathrm{g})^\mathrm{T}\mathbf{\Sigma}_\epsilon^{-1}(\mathbf{t-As}_\mathrm{g})+(\mathbf{s}_\mathrm{g}-\mathbf{s}_\mathrm{s})^\mathrm{T}\mathbf{\Sigma}_\mathrm{g}^{-1}(\mathbf{s}_\mathrm{g}-\mathbf{s}_\mathrm{s}) \\
 &\hspace{3ex}+\sum_i(\mathbf{D}\mathbf{x}_{i}-\mathbf{R}_i\mathbf{s}_\mathrm{s})^\mathrm{T}\mathbf{\Sigma}_{p,i}^{-1}(\mathbf{D}\mathbf{x}_{i}-\mathbf{R}_i\mathbf{s}_\mathrm{s})\bigg\} \\
 &\hspace{20ex} \text{subject to} \ \|\mathbf{x}_{i}\|_0=T \ \forall \ i.
% \label{eq:map2}
\end{split}\raisetag{2.5\baselineskip}
\end{equation}
%\end{linenomath*}

Further simplifying, we assume the errors are Gaussian iid. Thus, $\mathbf{\Sigma}_\epsilon=\sigma_\epsilon^2\mathbf{I}$, $\mathbf{\Sigma}_\mathrm{g}=\sigma_\mathrm{g}^2\mathbf{I}$, and $\mathbf{\Sigma}_{p,i}=\sigma_{p,i}^2\mathbf{I}$, where $\mathbf{I}$ is the identity matrix. The LST MAP objective is thus
%\begin{linenomath*}
\begin{equation}
\begin{split}
&\big\{\widehat{\mathbf{s}}_\mathrm{g},\widehat{\mathbf{s}}_\mathrm{s},\widehat{\mathbf{X}}\big\}= \underset{\mathbf{s}_\mathrm{g}, \mathbf{s}_\mathrm{s}, \mathbf{X}}{\arg\min}\ \bigg\{\frac{1}{\sigma_\epsilon^2}\|\mathbf{t-As}_\mathrm{g}\|_2^2 \\
 &\hspace{7ex}+\frac{1}{\sigma_\mathrm{g}^2}\|\mathbf{s}_\mathrm{g}-\mathbf{s}_\mathrm{s}\|_2^2+\frac{1}{\sigma_{p,i}^2}\sum_i\|\mathbf{D}\mathbf{x}_{i}-\mathbf{R}_i\mathbf{s}_\mathrm{s}\|_2^2\bigg\} \\ 
 &\hspace{15ex}\text{subject to} \ \|\mathbf{x}_{i}\|_0=T \ \forall \ i,
\label{eq:map6}
\end{split}
\end{equation}
%\end{linenomath*}
where $\big\{\widehat{\mathbf{s}}_\mathrm{g},\widehat{\mathbf{s}}_\mathrm{s},\widehat{\mathbf{X}}\big\}$ are the estimates of the LST variables.

%%%%%%%%%%%%%%%
\begin{figure*}
\centering
\includegraphics[width=18cm]{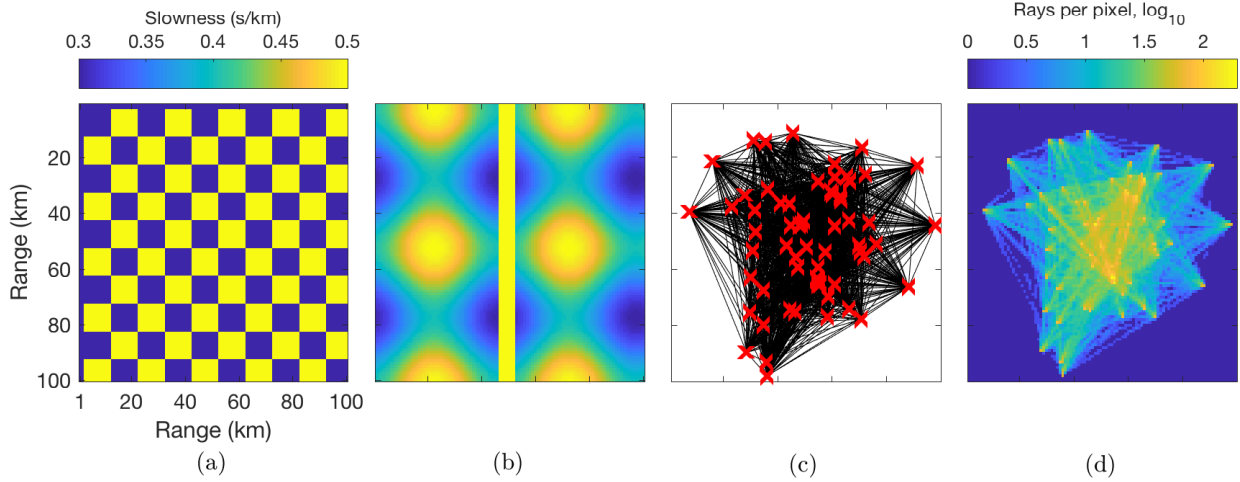}
\caption{Synthetic slowness maps and ray sampling. Slowness $\mathbf{s}'$ for (a) checkerboard map and (b) smooth-discontinuous map (sinusoidal variations with discontinuity). Both maps are $W_1=W_2=100$ pixels (1 km/pixel). (c) 64 stations (red X's), giving in 2016 straight ray (surface wave) paths through synthetic images. (d) Density of ray sampling, in $\log_{10}$ rays per pixel.}
\label{fig:mapsSampling}
\end{figure*}
%%%%%%%%%%%%%%%%
\subsection{Solving for the MAP estimate}
We find the MAP estimates $\big\{\widehat{\mathbf{s}}_\mathrm{g},\widehat{\mathbf{s}}_\mathrm{s},\widehat{\mathbf{X}}\big\}$ solving (\ref{eq:map6}) via a block-coordinate minimization algorithm \cite{elad2006,ravishankar2011}. This strategy divides the solution of (\ref{eq:map6}) into three subproblems: 1) the global problem corresponding, to the global model (\ref{eq:linearTraveltime}), which estimates $\widehat{\mathbf{s}}_\mathrm{g}$; 2) the local problem, corresponding to the local model (\ref{eq:patchForm1}) which estimates $\widehat{\mathbf{X}}$; and 3) an averaging procedure which estimates $\widehat{\mathbf{s}}_\mathrm{s}$. The global problem is formulated using least squares with $\ell_2$ regularization. This result is used to obtain local structure by solving for the patches from the sparse model (\ref{eq:patchForm1}), and averaging the patches. Global $\ell_2$ regularization has been used with TV regularization in seismic tomography\cite{lin2015}, which was an extension of \cite{huang2008}.  This modified-TV approach is adapted to travel time tomography as a competing method in Section~\ref{sec:tv}. LST combines a global $\ell_2$-norm constraint with local regularization of patches, following \cite{elad2010}. A major distinction between LST and image denoising \cite{elad2006} is that the 2D image is inferred from travel time measurements, and not directly observed, similar to \cite{ravishankar2011}.

The global problem is written from (\ref{eq:map6}) as
%\begin{linenomath*}
\begin{equation}
\begin{aligned}
\widehat{\mathbf{s}}_\mathrm{g}= & \underset{\mathbf{s}_\mathrm{g}}{\arg\min}\ \frac{1}{\sigma_\epsilon^2}\|\mathbf{t-As}_\mathrm{g}\|_2^2+\frac{1}{\sigma_\mathrm{g}^2}\|\mathbf{s}_\mathrm{g}-\mathbf{s}_\mathrm{s}\|_2^2 \\
= &\underset{\mathbf{s}_\mathrm{g}}{\arg\min}\ \|\mathbf{t-As}_\mathrm{g}\|_2^2+\lambda_1\|\mathbf{s}_\mathrm{g}-\mathbf{s}_\mathrm{s}\|_2^2,
\label{eq:mapGlobal2}
\end{aligned}
\end{equation}
%\end{linenomath*}
where $\lambda_1=(\sigma_\epsilon/\sigma_\mathrm{g})^2$ is a regularization parameter.

The local problem is written from (\ref{eq:map6}), with each patch solved from the global estimate $\mathbf{s}_\mathrm{s}=\widehat{\mathbf{s}}_\mathrm{g}$ from (\ref{eq:mapGlobal2})  (decoupling the local and global problems), giving
%\begin{linenomath*}
\begin{equation}
\widehat{\mathbf{x}}_i= \underset{\mathbf{x}_i}{\arg\min} \ \|\mathbf{D}\mathbf{x}_{i}-\mathbf{R}_i\widehat{\mathbf{s}}_\mathrm{g}\|_2^2 \ \ \text{subject to} \ \|\mathbf{x}_{i}\|_0=T.
\label{eq:mapLocal1}
\end{equation}
%\end{linenomath*}
With the estimate of coefficients $\widehat{\mathbf{X}}=[\widehat{\mathbf{x}}_1,...,\widehat{\mathbf{x}}_I]$ from (\ref{eq:mapLocal1}) and global slowness $\widehat{\mathbf{s}}_\mathrm{g}$ from (\ref{eq:mapGlobal2}) we solve for $\mathbf{s}_\mathrm{s}$. (\ref{eq:map6}) gives, assuming constant patch variance $\sigma_{p,i}^2=\sigma_{p}^2$,
%\begin{linenomath*}
\begin{equation}
\begin{aligned}
\widehat{\mathbf{s}}_\mathrm{s}= &\underset{\mathbf{s}_\mathrm{s}}{\arg\min}\ \frac{1}{\sigma_\mathrm{g}^2}\|\mathbf{\widehat{s}}_\mathrm{g}-\mathbf{s}_\mathrm{s}\|_2^2+\frac{1}{\sigma_{p}^2}\sum_i\|\mathbf{D}\widehat{\mathbf{x}}_{i}-\mathbf{R}_i\mathbf{s}_\mathrm{s}\|_2^2 \\
= &\underset{\mathbf{s}_s}{\arg\min}\ \lambda_2\|\mathbf{\widehat{s}}_\mathrm{g}-\mathbf{s}_\mathrm{s}\|_2^2+\sum_i\|\mathbf{D}\widehat{\mathbf{x}}_{i}-\mathbf{R}_i\mathbf{s}_\mathrm{s}\|_2^2,
\label{eq:mapLocal4}
\end{aligned}
\end{equation}
%\end{linenomath*}
where $\lambda_2=(\sigma_p/\sigma_\mathrm{g})^2$ is a regularization parameter. The solution to (\ref{eq:mapLocal4}) is analytic, with $\widehat{\mathbf{s}}_\mathrm{s}$ the stationary point. Differentiating (\ref{eq:mapLocal4}) gives
%\begin{linenomath*}
\begin{equation}
\begin{aligned}
\frac{d}{d\mathbf{s}_\mathrm{s}}&\bigg\{\lambda_2\|\widehat{\mathbf{s}}_\mathrm{g}-\mathbf{s}_\mathrm{s}\|_2^2+\sum_i\|\mathbf{D}\widehat{\mathbf{x}}_{i}-\mathbf{R}_i\mathbf{s}_\mathrm{s}\|_2^2\bigg\} \\
=&\lambda_2\big(\mathbf{s}_\mathrm{s}-\widehat{\mathbf{s}}_\mathrm{g}\big)+\sum_i\mathbf{R}_i^\mathrm{T}\big(\mathbf{R}_i\mathbf{s}_\mathrm{s}-\mathbf{D}\widehat{\mathbf{x}}_{i}\big)\\
=&\big(\lambda_2\mathbf{I}+\sum_i\mathbf{R}_i^\mathrm{T}\mathbf{R}_i\big)\mathbf{s}_\mathrm{s}-\lambda_2\widehat{\mathbf{s}}_\mathrm{g}-\sum_i\mathbf{R}_i^\mathrm{T}\mathbf{D}\widehat{\mathbf{x}}_{i} \\
=&\big(\lambda_2\mathbf{I}+n\mathbf{I}\big)\mathbf{s}_\mathrm{s}-\lambda_2\widehat{\mathbf{s}}_\mathrm{g}-\mathbf{s}_p,
\label{eq:mapLocal6}
\end{aligned}
\end{equation}
%\end{linenomath*}
where $n\mathbf{I}=\sum_i\mathbf{R}_i^\mathrm{T}\mathbf{R}_i$ and $\mathbf{s}_p=\frac{1}{n}\sum_i\mathbf{R}_i^\mathrm{T}\mathbf{D}\widehat{\mathbf{x}}_{i}$. Thus we obtain
%%%%%%%%%%%%%%%
%\begin{linenomath*}
\begin{equation}
\widehat{\mathbf{s}}_{\mathrm{s}}=\frac{\lambda_2\widehat{\mathbf{s}}_{\mathrm{g}}+n\mathbf{s}_{p}}{\lambda_2+n},
\label{eq:mapLocal8}
\end{equation}
%\end{linenomath*}
%%%%%%%%%%%%%%%%
which gives $\mathbf{s}_\mathrm{s}$ as the weighted average of the patch slownesses $\{\mathbf{D}\widehat{\mathbf{x}}_{i}\ \forall\ i \}$ and $\widehat{\mathbf{s}}_\mathrm{g}$. When $\lambda_2\ll n$, $\mathbf{s}_\mathrm{s}\approx\mathbf{s}_p$. When $\lambda_2=n$, $\mathbf{s}_\mathrm{g}$ and $\mathbf{s}_p$ have equal weight. It is typical in image denoising to set $\lambda_2=0$ \cite{mairal2014}.
 
\subsection{LST algorithm with dictionary learning}
\label{sec:st3dlearn}
The expressions (\ref{eq:mapGlobal2}), (\ref{eq:mapLocal1}), and (\ref{eq:mapLocal8}) are solved iteratively, giving the LST algorithm as a MAP estimate of a slowness image with local sparse constraints with a known dictionary $\mathbf{D}$. Dictionary learning is added to the algorithm in the solution to the local problem (\ref{eq:mapLocal1}), by optimizing $\mathbf{D}$:
%\begin{linenomath*}
\begin{equation}
\begin{aligned}
\widehat{\mathbf{D}}= \underset{\mathbf{D}}{\arg\min}&\big\{\underset{\mathbf{x}_i}{\min} \ \|\mathbf{D}\mathbf{x}_{i}-\mathbf{R}_i\widehat{\mathbf{s}}_\mathrm{g}\|_2^2 \ \\
& \text{subject to} \ \|\mathbf{x}_{i}\|_0=T~\forall~i\big\}.
\label{eq:mapLocal1_aug}
\end{aligned}
\end{equation}
%\end{linenomath*}
The dictionary learning problem (\ref{eq:mapLocal1_aug}) is here solved using the ITKM algorithm, Table \ref{algo:itkm} (for details, see App.~\ref{sec:appendixITKM}). 

The ITKM solves this bilinear optimization problem (\ref{eq:mapLocal1_aug}) by alternately solving for the sparse coefficients $\widehat{\mathbf{x}}_i$ with $\mathbf{D}$ fixed, and solving for $\widehat{\mathbf{D}}$ with $\widehat{\mathbf{x}}_i$ fixed. The columns of $\mathbf{D}$, $\mathbf{d}_q$, are constrained to have unit norm, to prevent scaling ambiguity. $\widehat{\mathbf{x}}_i$ are solved using thresholding, and $\mathbf{D}$ is estimated using a `signed' K-means objective. After the dictionary is obtained from ITKM, $\mathbf{x}_i$ is solved OMP for all patches. For fixed sparsity $T$, the ITKM \cite{schnass2015} is more computationally efficient and has better guarantees of dictionary recovery than the K-SVD\cite{aharon2006}.

\begin{figure}[t]
\centering
\includegraphics[width=7cm]{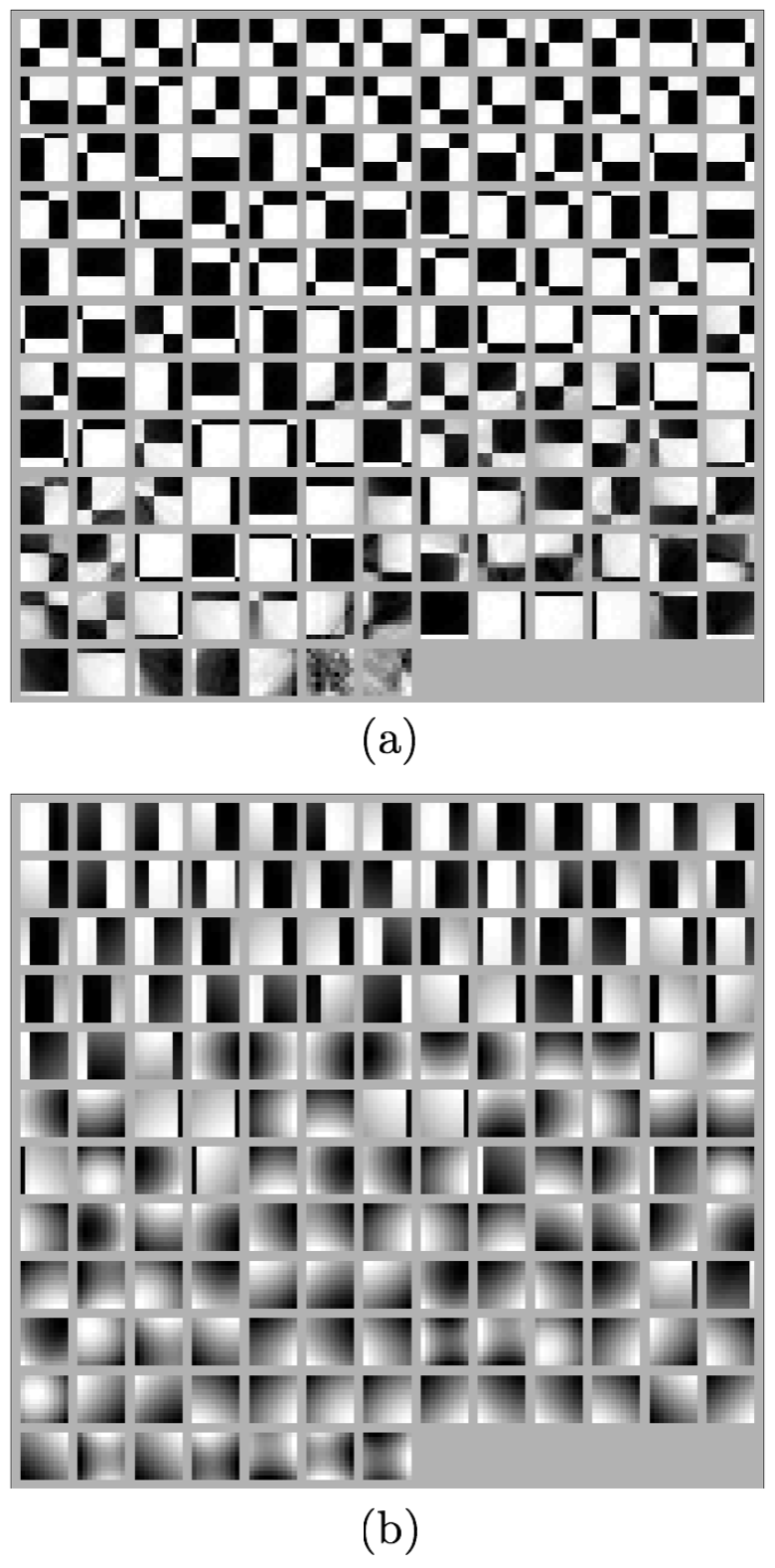} \vspace{1ex}
\caption{Dictionary atoms learned during LST with ITKM, with $n=100$ and $Q=150$ for (a) checkerboard map (Fig. \ref{fig:mapsSampling}(a)) with $T=1$ and (b) smooth-discontinuous map (Fig. \ref{fig:mapsSampling}(b)) with $T=2$. Atoms adjusted to full grayscale range for display.} 
\label{fig:learnedDicts}
\end{figure}

To illustrate the content of learned dictionaries, the atoms learned during LST are shown for the checkerboard (Fig.~\ref{fig:learnedDicts}(a)) and smooth-discontinuous map (Fig.~\ref{fig:learnedDicts}(b)). The atoms from checkerboard (Fig. \ref{fig:learnedDicts}(a)) contain sharp edges, which correspond to shifted sharp edges from the checkerboard pattern. The atoms from smooth-discontinuous map (Fig. \ref{fig:learnedDicts}(b)) contain both smooth and discontinuous features. The smooth atoms correspond to the sinusoidal variations, whereas the atoms with sharp edges correspond to shifted features the fault region. These features give the \textit{shift-invariance} property to the sparse, representation, which will be discussed in Sec.~\ref{sec:advan}. Since learned dictionaries are adapted to specific data, they better model specific data with a minimal number of atoms than prescribed dictionaries, such as Haar wavelets or DCT. Methods using dictionary learning have obtained superior performance over prescribed dictionaries in e.g. image denoising and inpainting\cite{mairal2014}.

Since $\mathbf{A}$ is sparse, the global estimate (\ref{eq:mapGlobal2}) is solved using the sparse least squares program LSQR \cite{paige1982}. The local estimate (\ref{eq:mapLocal1}) is solved using orthogonal matching pursuit (OMP) \cite{pati93} after the slowness patches $\{\mathbf{R}_i\widehat{\mathbf{s}}_\mathrm{g} \ \forall \ i \ \}$ are centered \cite{mairal2014} -- i.e. the mean of the pixels in each patch is subtracted. The mean of patch $i$ is $\overline{x}_{i}=\frac{1}{n}\mathbf{1}^\mathrm{T}\mathbf{R}_i\widehat{\mathbf{s}}_\mathrm{g}$. Hence, $\mathbf{R}_i\widehat{\mathbf{s}}_\mathrm{g}\approx\mathbf{Dx}_i+\mathbf{1}\overline{x}_i$. The algorithm for the MAP estimate with and without dictionary learning is given in Table \ref{algo:patchSparse}.

The complexity of each LST iteration is determined primarily by LSQR computation in the global problem, $O(2MN)$, and in the local problem by ITKM $O(knQN)$ and OMP $O(TnQN)$, where $k$ is the ITKM iterations (see Table \ref{algo:patchSparse}). For large slowness maps, we expect the LST complexity to be dominated by LSQR. In our simulations we obtain reasonable run times (see Sec.~\ref{sec:convergence}). In the special case when $T=1$, the solution to (\ref{eq:mapLocal1}) is not combinatorial, and the dictionary learning problem is equivalent gain-shape vector quantization\cite{aharon2006,gersho1991}.

\subsection{Advantages}
\label{sec:advan}
Improved inversion performance over conventional TV regularized tomography is obtained under the hypothesis that the seismic image patches can be represented as sparse linear combinations of a small set of elemental patches, or patterns. Such patterns are the atoms in the dictionary $\mathbf{D}$. This hypothesis follows numerous works in image processing and neuroscience communities, for example \cite{olshausen1997,mairal2009,elad2010,hyvarinen2014}, which have shown that patches of diverse image content are well approximated with a sparse linear combinations of atoms. This property has been exploited for signal denoising and inpainting \cite{elad2010,mairal2009} and classification \cite{mairal2012,fedorov2017}. 

As previously mentioned, sparse dictionaries can model diverse image content. Further, sparse dictionaries trained on overlapping patches possess the shift-invariance property, whereby features such as edges are recovered regardless of where they are located in an image \cite{elad2010}. LST enables finer resolution as permitted by the atoms in the dictionary and can exploit shift invariance. Slowness features in Fig. \ref{fig:mapsSampling}(a--b) are shifted such that only small cells of constant slowness may be used with conventional tomography (which necessitates damping, Sec.\ \ref{sec:conventional}) to illustrate this effect.

%%%%%%%%%%%%%%%%
%%%%%%%%%%%%%%%%
 \begin{table}[t!]
\begin{center}
\caption{Locally-sparse travel time tomography (LST) algorithm}
\begin{tabular}{ll}
\hline\hline
Given: $\mathbf{t}\in\mathbb{R}^{K}$, $\mathbf{A}\in\mathbb{R}^{M\times N}$, $\mathbf{s}_\mathrm{s}^0=\mathbf{0}\in\mathbb{R}^N$, $\mathbf{D}^0= \text{Haar wavelet,}$ \\ 
 \hspace{2ex}$\text{DCT (or) noise}\ \mathcal{N}\big(0,1\big)\in\mathbb{R}^{n\times Q}$, $\lambda_1$, $\lambda_2$, $T$, and $j=1$ \\ \hline
  Repeat until convergence: \\
  1. \text{Global estimate}: solve (\ref{eq:mapGlobal2}) using LSQR\cite{paige1982}, \\
   \hspace{3ex} $\widehat{\mathbf{s}}_\mathrm{g}^j=\underset{\mathbf{s}_\mathrm{g}^j}{\arg\min}\ \|\mathbf{As}_\mathrm{g}^j-\mathbf{t}\|_2^2 +\lambda_1 \|\mathbf{s}_\mathrm{g}^j-\mathbf{s}_\mathrm{s}^{j-1}\|_2^2$. \\
  2. \text{Local estimate}\\
\ \ a: \ Setting $\mathbf{s}_\mathrm{s}^j=\widehat{\mathbf{s}}_\mathrm{g}^j$, center patches $\{\mathbf{R}_i\widehat{\mathbf{s}}_\mathrm{g}^j \ \forall \ i \ \}$ and \\
\hspace{3ex} i. (Dictionary learning) Solve (\ref{eq:mapLocal1_aug}) for $\mathbf{D}^j$ using ITKM \cite{schnass2015} (Table \ref{algo:itkm}), \\
\hspace{1ex}  $\widehat{\mathbf{D}}^j= \underset{\mathbf{D}^j}{\arg\min}\big\{\underset{\mathbf{x}_i}{\min} \ \|\mathbf{D}^{j-1}\mathbf{x}_{i}-\mathbf{R}_i\widehat{\mathbf{s}}_\mathrm{g}\|_2^2~\text{subject to} \ \|\mathbf{x}_{i}\|_0=T~\forall~i\big\}$ \\
\hspace{3ex} i. (generic dictionary)  Set $\mathbf{D}^j=\mathbf{D}^0$. \\
\hspace{3ex} ii. Solve (\ref{eq:mapLocal1}) using OMP, \\
\hspace{1ex}  $\widehat{\mathbf{x}}_i^j= \underset{\mathbf{x}_i^j}{\arg\min} \ \|\mathbf{D}^j\mathbf{x}_{i}^j-\mathbf{R}_i\widehat{\mathbf{s}}_\mathrm{g}^j\|_2^2~\text{subject to} \ \|\mathbf{x}_{i}^j\|_0=T~\forall~i$. \\
\ \ b: Obtain $\widehat{\mathbf{s}}_\mathrm{s}^j$ by (\ref{eq:mapLocal8}) 
$\widehat{\mathbf{s}}_{\mathrm{s}}^j=\dfrac{\lambda_2\widehat{\mathbf{s}}_{\mathrm{g}}^j+n\mathbf{s}_{\mathrm{p}}^j}{\lambda_2+n}$ \\
\hspace{2ex} $j=j+1$ \\
\hline\hline
\end{tabular}
\label{algo:patchSparse}
\end{center}
\end{table}
%%%%%%%%%%%

\begin{figure}
\centering
\includegraphics[width=6cm]{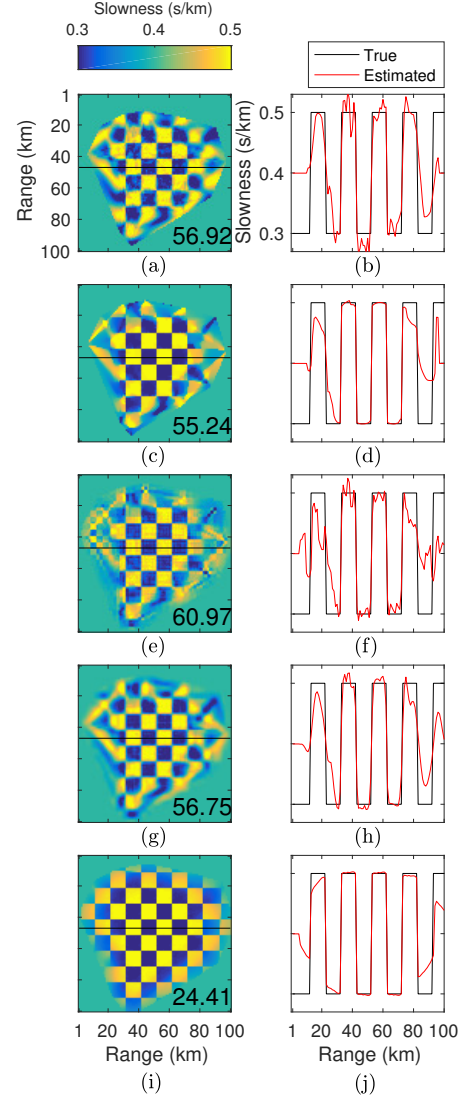}
\caption{Conventional, TV, and LST tomography results without travel time error for checkerboard map (Fig. \ref{fig:mapsSampling}(a)): 2D and 1D (from black line in 2D) slowness estimates plotted against true slowness $\mathbf{s}'$. (a,b) conventional, (c,d) TV regularization, $\widehat{\mathbf{s}}_\text{TV}'$, and LST with (e,f) Haar wavelet and (g,h) DCT dictionary with $Q=169$, $T=5$, $n=64$; and (g,h) with dictionary learning. RMSE (ms/km), per (\ref{eq:rmse}), is printed on 2D slownesses.}
\label{fig:checker_noNoise}
\end{figure}

\section{Competing methods}
\subsection{Conventional tomography}
\label{sec:conventional}
We illustrate conventional tomography with a Bayesian approach \cite{rodgers2000}, which enforces smoothness regularization with a global (non-diagonal) covariance. Considering the measurements (\ref{eq:linearTraveltime}), the MAP estimate of the slowness is
%\begin{linenomath*}
\begin{equation}
\widehat{\mathbf{s}}_\mathrm{g}=\big(\mathbf{A}^\mathrm{T}\mathbf{A}+\eta\mathbf{\Sigma}_L^{-1}\big)^{-1}\mathbf{A}^\mathrm{T}\mathbf{t},
\label{eq:classicMAP_solution1}
\end{equation}
%\end{linenomath*}
where $\eta=(\sigma_\epsilon/\sigma_\mathrm{c})^2$ is a regularization parameter, $\sigma_\mathrm{c}$ is the conventional slowness variance, and
%\begin{linenomath*}
\begin{equation}
\mathbf{\Sigma}_L(i,j)=\exp\big(-D_{i,j}/L\big).
\label{eq:classicMAP_solution2}
\end{equation}
%\end{linenomath*}
Here $D_{i,j}$ is the distance between cells $i$ and $j$, and $L$ is the smoothness length scale \cite{rodgers2000, tarantola1987}.

\subsection{Total variation regularization}
\label{sec:tv}
We implement the modified TV regularization method \cite{huang2008,lin2015}. TV regularization penalizes the gradient between pixels, enforcing piecewise constant models \cite{chambolle2004}, hence we might expect TV regularization to preserve well discontinuous or constant features. The TV method is adapted to the travel time tomography problem, giving the objective 
\begin{gather}
\begin{aligned}
\big\{\widehat{\mathbf{s}}_\mathrm{g},\widehat{\mathbf{s}}_\mathrm{TV}\big\}= \underset{\mathbf{s}_\mathrm{g}, \mathbf{s}_\mathrm{TV}}{\arg\min}&\ \bigg\{\frac{1}{\sigma_\epsilon^2}\|\mathbf{t-As}_\mathrm{g}\|_2^2 \\
 &+\frac{1}{\sigma_\mathrm{g}^2}\|\mathbf{s}_\mathrm{g}-\mathbf{s}_\mathrm{TV}\|_2^2+\mu\|\mathbf{s}_\mathrm{TV}\|_\mathrm{TV}\bigg\},
 \label{eq:tv1} \raisetag{3\baselineskip}
\end{aligned}
\end{gather}
where $\|\mathbf{s}_\mathrm{TV}\|_\mathrm{TV}$ is the TV regularizer, which penalizes the gradient, and $\mathbf{s}_\mathrm{TV}\in\mathbb{R}^N$ is the TV estimate of the slowness. Similar to LST, TV (\ref{eq:tv1}) is solved by decoupling the problem into two subproblems: 1) damped least squares and 2) TV. The least squares problem is
\begin{equation}
\begin{split}
\widehat{\mathbf{s}}_\mathrm{g}=& \underset{\mathbf{s}_\mathrm{g}}{\arg\min}\ \frac{1}{\sigma_\epsilon^2}\|\mathbf{t-As}_\mathrm{g}\|_2^2+\frac{1}{\sigma_\mathrm{g}^2}\|\mathbf{s}_\mathrm{g}-\mathbf{s}_\mathrm{TV}\|_2^2 \\
=&\underset{\mathbf{s}_\mathrm{g}}{\arg\min}~\|\mathbf{t-As}_\mathrm{g}\|_2^2+\lambda_1\|\mathbf{s}_\mathrm{g}-\mathbf{s}_\mathrm{TV}\|_2^2,
 \label{eq:tv2}
\end{split}
\end{equation}
where $\lambda_1$ is related to global LST problem (see (\ref{eq:mapGlobal2})) and $\mathbf{s}_\mathrm{TV}$ is initialized to the reference slowness. The TV problem is
\begin{equation}
\begin{split}
\widehat{\mathbf{s}}_\mathrm{TV}=& \underset{\mathbf{s}_\mathrm{TV}}{\arg\min}~\frac{1}{\sigma_\mathrm{g}^2}\|\mathbf{s}_\mathrm{g}-\mathbf{s}_\mathrm{TV}\|_2^2+\mu\|\mathbf{s}_\mathrm{TV}\|_\mathrm{TV} \\
=&\underset{\mathbf{s}_\mathrm{TV}}{\arg\min}~\|\mathbf{s}_\mathrm{g}-\mathbf{s}_\mathrm{TV}\|_2^2+\lambda_\mathrm{TV}\|\mathbf{s}_\mathrm{TV}\|_\mathrm{TV},
 \label{eq:tv3}
\end{split}
\end{equation}
with $\lambda_\mathrm{TV}=\sigma_\mathrm{g}^2\mu$. (\ref{eq:tv1}) is solved by alternately solving (\ref{eq:tv2}) and (\ref{eq:tv3}), like LST, as a block coordinate minimization algorithm. In this paper (\ref{eq:tv2}) is solved using LSQR \cite{paige1982} and (\ref{eq:tv3}) is solved using the TV algorithm of Chambolle \cite{chambolle2004,zhu2010}. We set the gradient step size $\alpha=0.25$, which is optimal for convergence and stability of the algorithm \cite{chambolle2004}. The stopping tolerance is set to 1e-2.

\section{Simulations}
\begin{figure}
\centering
\includegraphics[width=9cm]{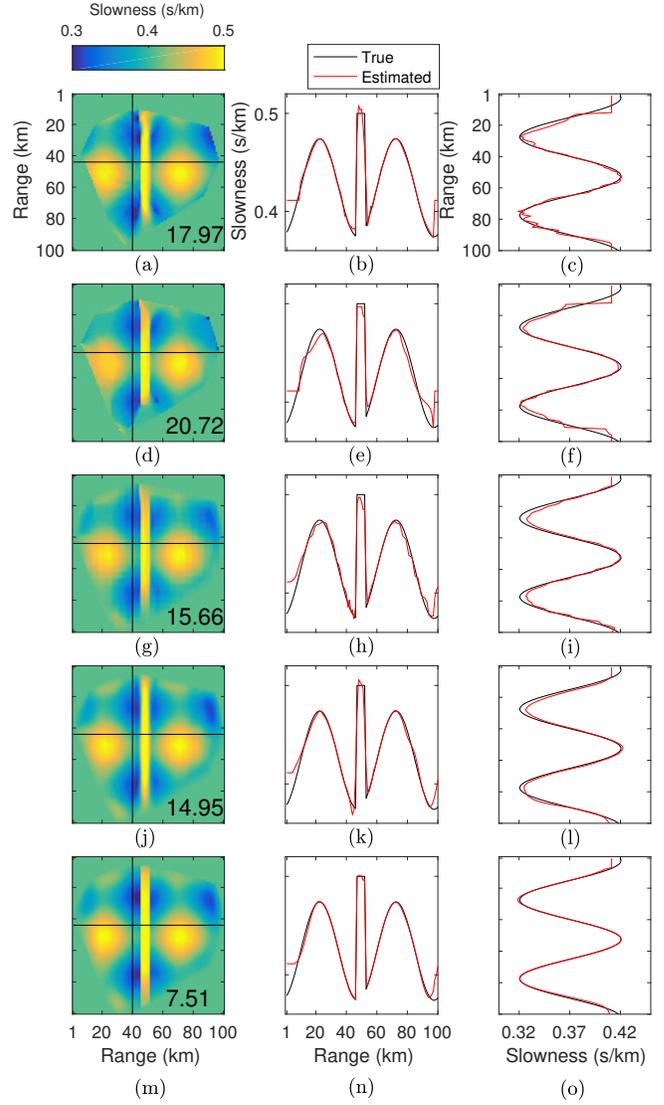}
\caption{Conventional, TV, and LST tomography results without travel time error for smooth-discontinuous map (Fig. \ref{fig:mapsSampling}(b)). Same order as Fig.~5, with vertical 1D profile of estimated and true slowness. RMSE (ms/km), per (\ref{eq:rmse}), is printed on 2D slownesses.}
\label{fig:smoothDiscon_noNoise}
\end{figure}

\begin{figure*}[!t]
\centering
\subfloat{\includegraphics[width=6cm]{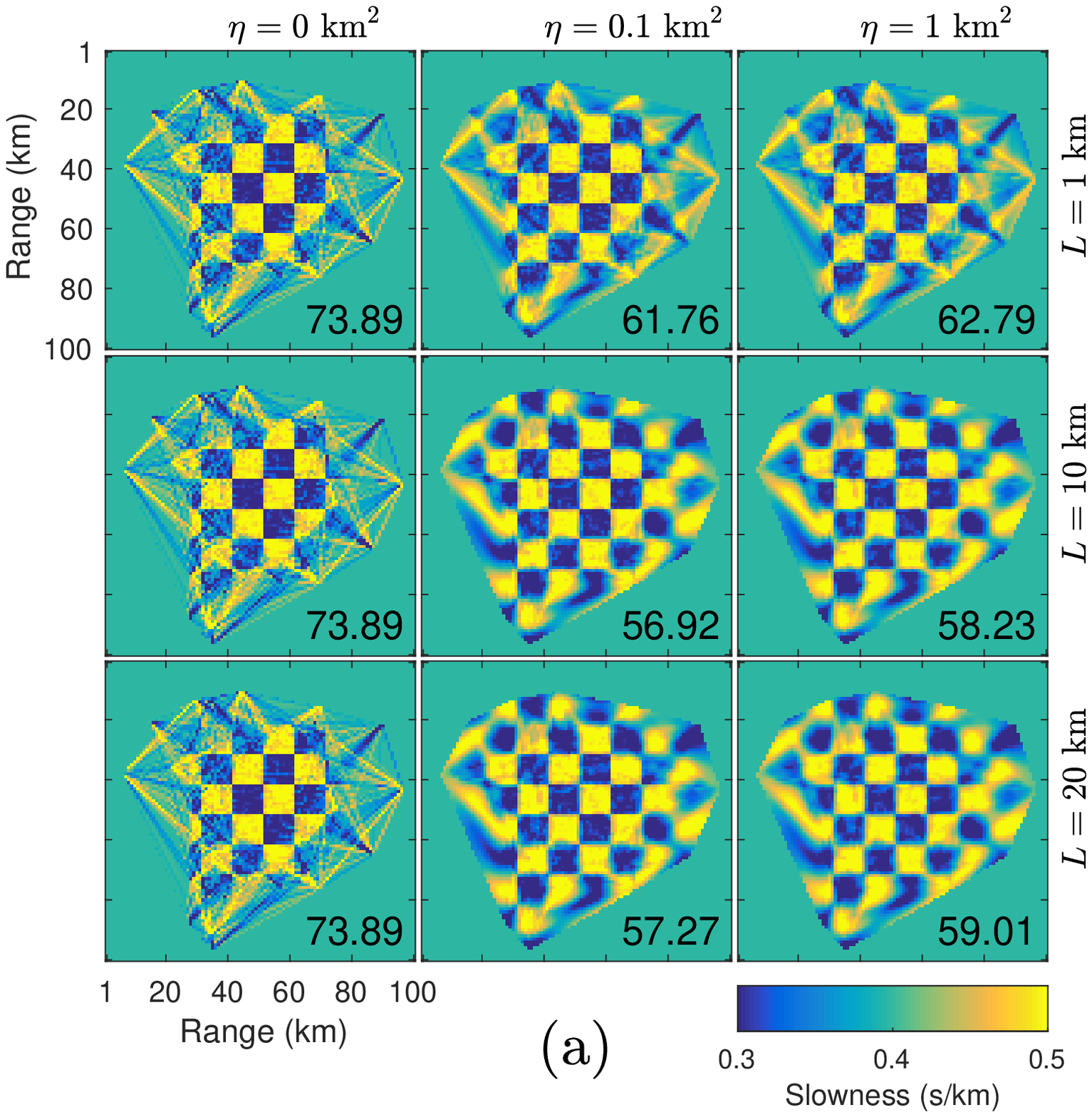}\hspace{-.2cm}
\label{fig:noNoise_conv}}
\subfloat{\includegraphics[width=6cm]{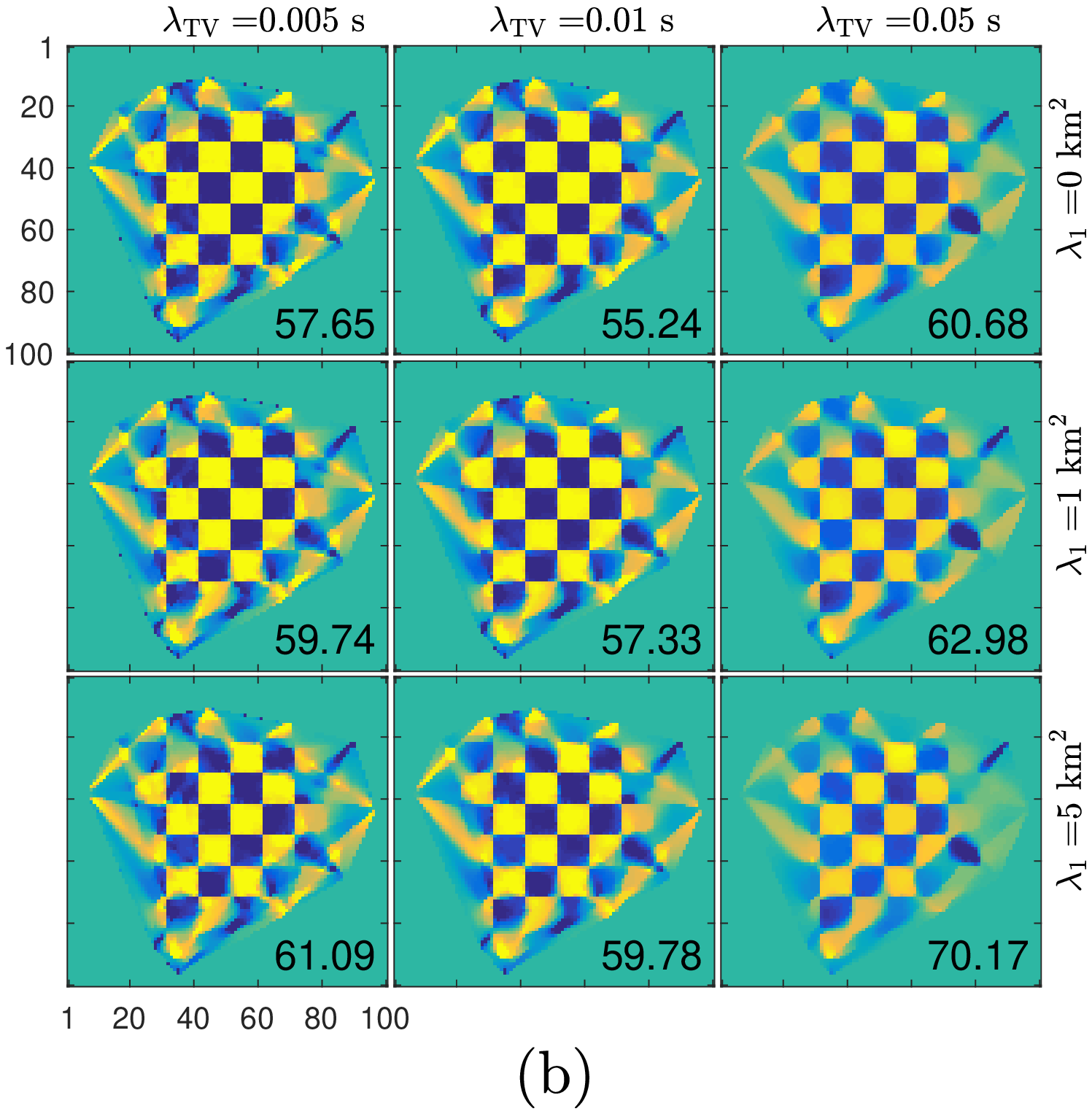}\hspace{-.2cm}
\label{fig_second_case}}
\subfloat{\includegraphics[width=6cm]{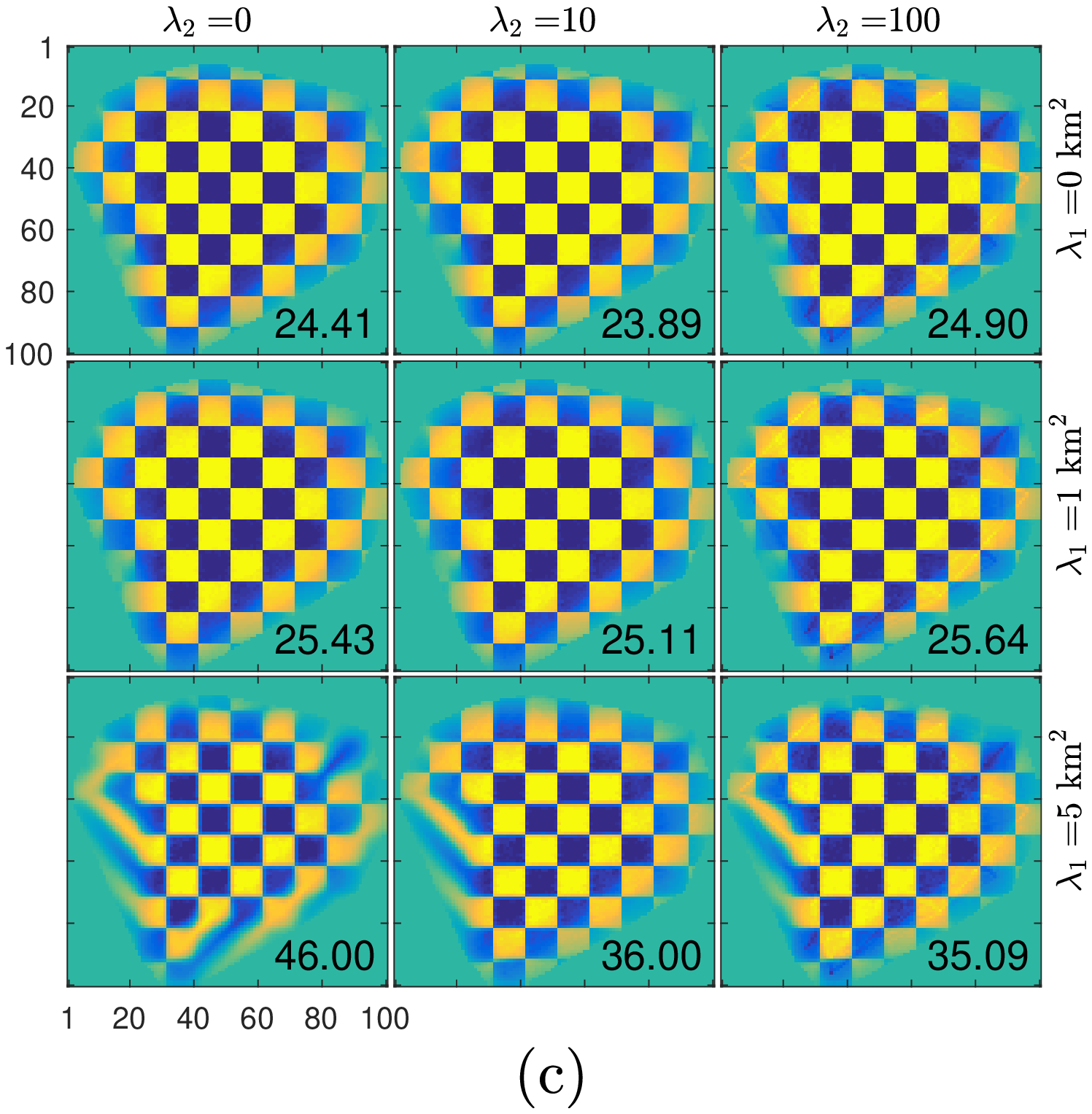}
\label{fig_second_case}}\vspace{-.5cm}
\subfloat{\includegraphics[width=6cm]{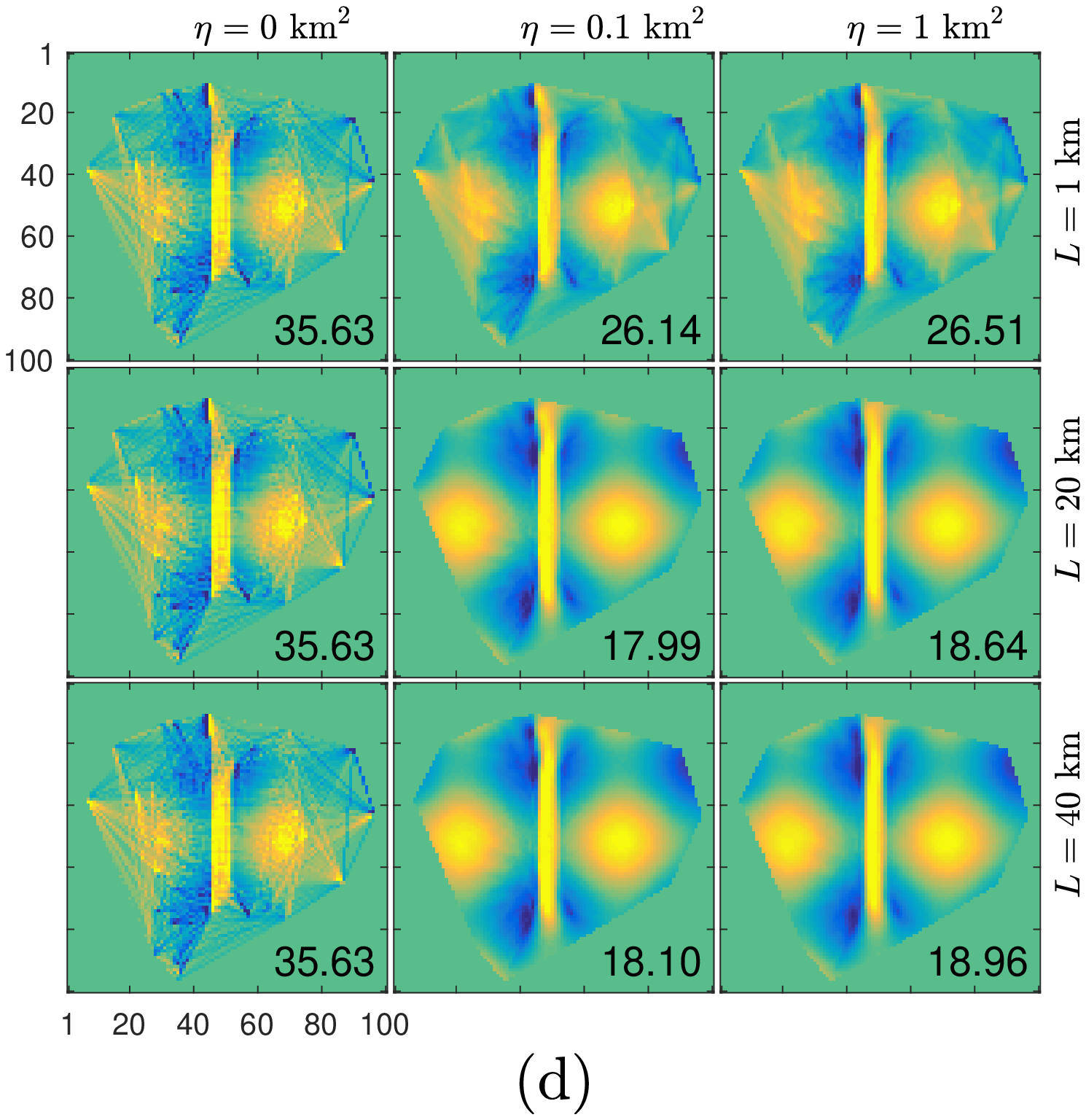}\hspace{-.2cm}
\label{fig_first_case}}
\subfloat{\includegraphics[width=6cm]{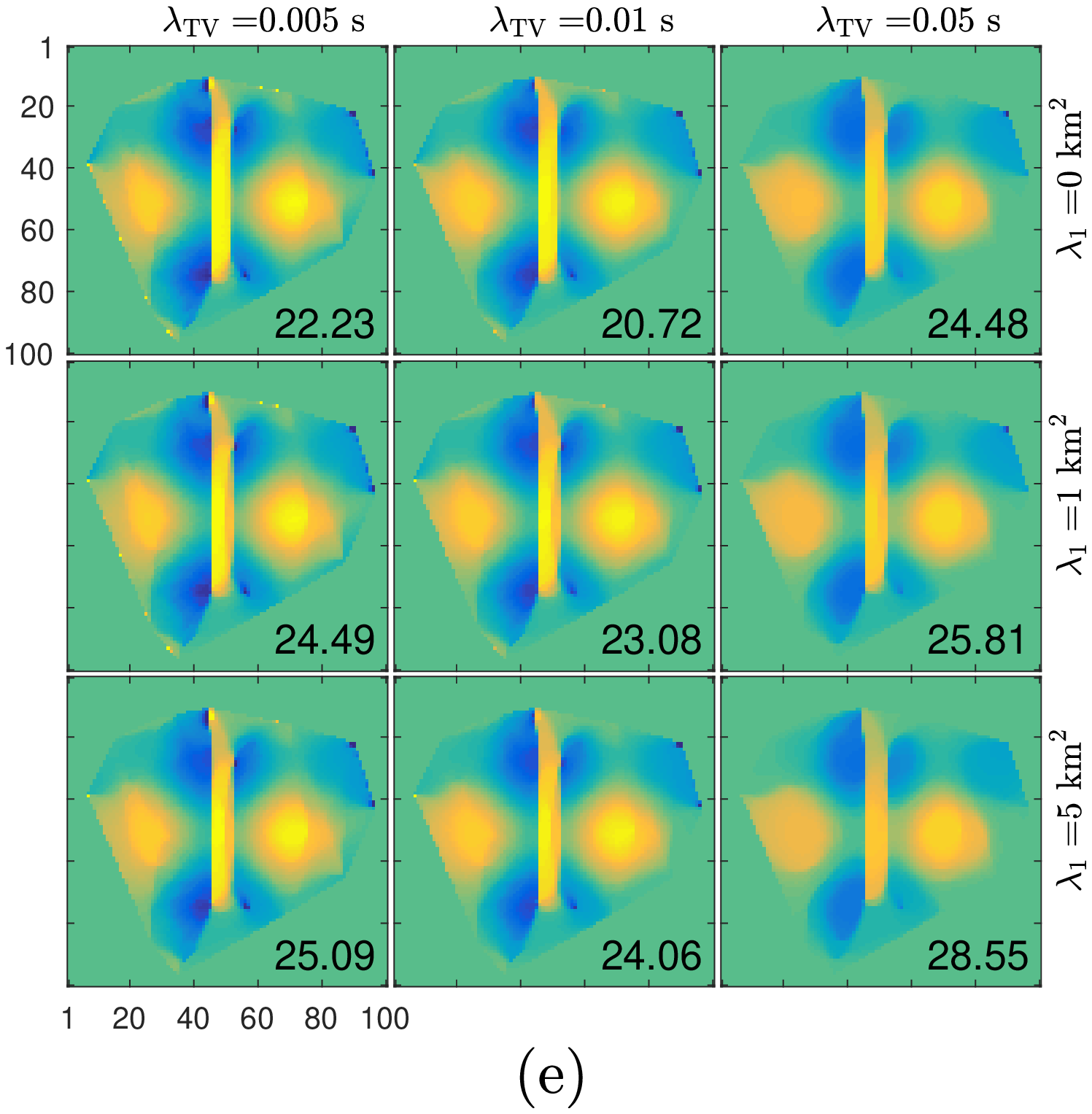}\hspace{-.2cm}
\label{fig_second_case}}
\subfloat{\includegraphics[width=6cm]{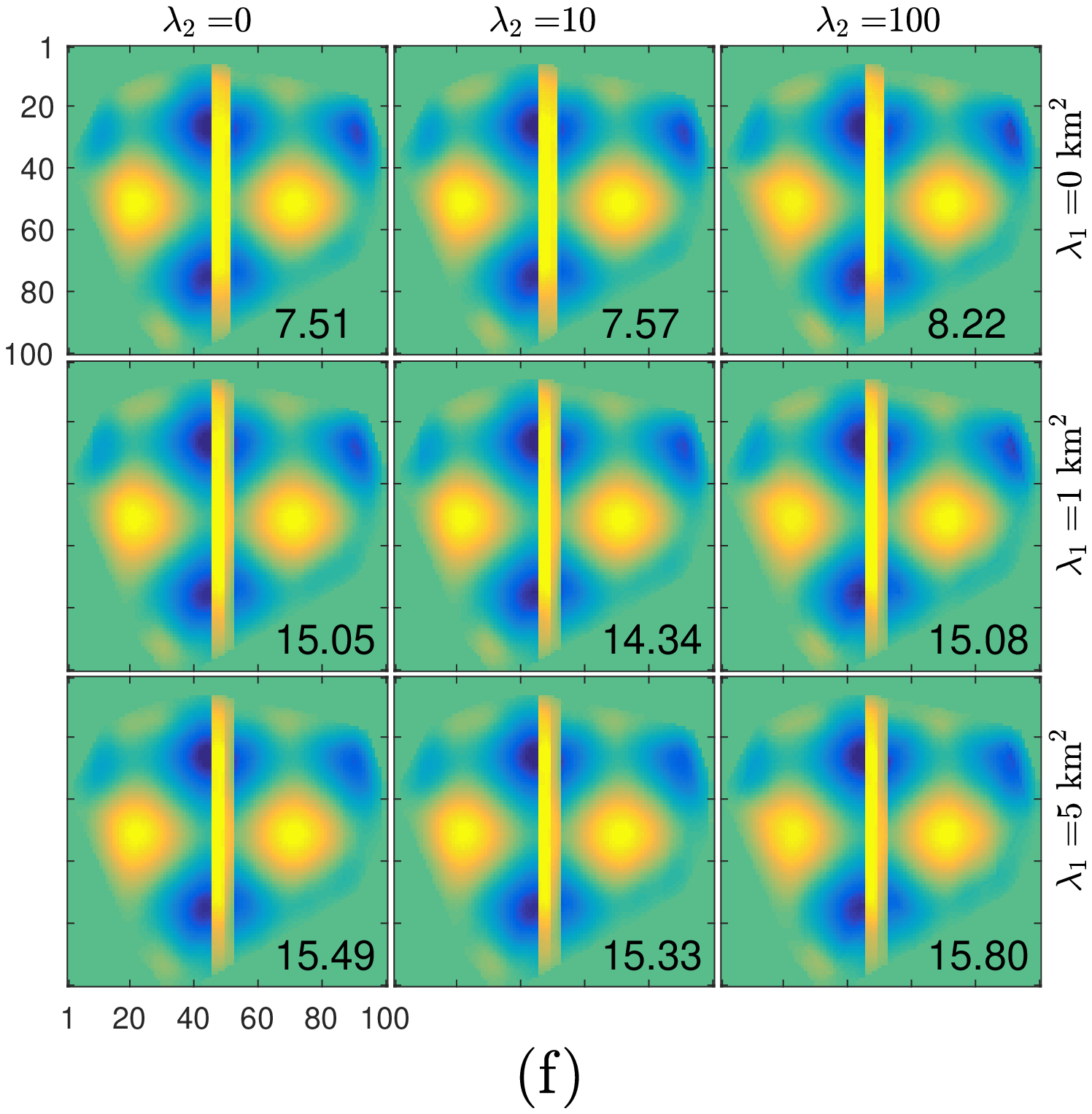}
\label{fig_second_case}}
\caption{Conventional, TV, and LST tomography results for different values of regularization parameters for (a--c) checkerboard (Fig. \ref{fig:mapsSampling}(a)) and (d--f) smooth-discontinuous (Fig. \ref{fig:mapsSampling}(b)) maps, without travel time error. (a,d) Conventional $\widehat{\mathbf{s}}'_\mathrm{g}$, effect of $L$ and $\eta$. (b,e) TV regularization $\widehat{\mathbf{s}}'_\text{TV}$, effect of $\lambda_1$ and $\lambda_\text{TV}$. (c,f) LST $\widehat{\mathbf{s}}'_\mathrm{s}$, effect of $\lambda_1$ and $\lambda_2$. RMSE (ms/km), per (\ref{eq:rmse}), is printed on 2D slownesses.}
\label{fig:noNoise_parmVary}
\end{figure*}

We demonstrate the performance of LST (Sec.~\ref{sec:propMAP}, Table~\ref{algo:patchSparse}), using both dictionary learning and prescribed dictionaries, relative to conventional (Sec.~\ref{sec:conventional}) and TV (Sec.~\ref{sec:tv}) tomography on synthetic slowness maps (e.g. Fig.~\ref{fig:mapsSampling}(a,b)). The recovered slownesses from the methods are plotted in Figs.~\ref{fig:checker_noNoise}--\ref{fig:mc}, and their performance is summarized in Table \ref{table:rmse}. The convergence of LST and its sensitivity to sparsity $T$ are shown in Fig.~\ref{fig:errorCurves}. 

Experiments are conducted using simulated seismic data from two synthetic 2D seismic slowness maps (Fig. \ref{fig:mapsSampling}(a,b)) with dimensions $W_1=W_2=100$ pixels (km), as well as variations of these maps. The boxcar checkerboard in Fig.~\ref{fig:mapsSampling}(a) demonstrates the recovery of discontinuous seismic slownesses. While the checkerboard slowness is quite unrealistic, it is commonly used as a benchmark for seismic tomography methods. The smooth-discontinuous map Fig.~\ref{fig:mapsSampling}(b), is more realistic and illustrates fault-like discontinuities in an otherwise smoothly varying (sinusoidal) slowness map, as used in \cite{loris2007}. These examples illustrate the modeling flexibility of the LST algorithm. We also generate a variety of synthetic slowness maps, by altering the size of the checkerboard squares and the width and horizontal location of the discontinuity, in the checkerboard and smooth-discontinuous slowness maps, respectively. For more details, see Sec.~\ref{sec:noiseFree_results}. The travel times from the synthetic slowness maps are generated by the global model (\ref{eq:linearTraveltime}). The slowness estimates from LST are $\widehat{\mathbf{s}}_\mathrm{s}'=\widehat{\mathbf{s}}_\mathrm{s}+\mathbf{s}_0\in\mathbb{R}^N$ (from (\ref{eq:mapLocal8})), for conventional $\widehat{\mathbf{s}}_\mathrm{g}'=\widehat{\mathbf{s}}_\mathrm{g}+\mathbf{s}_0\in\mathbb{R}^N$ (from (\ref{eq:classicMAP_solution1})), and for TV $\widehat{\mathbf{s}}_\mathrm{TV}'=\widehat{\mathbf{s}}_\mathrm{TV}+\mathbf{s}_0\in\mathbb{R}^N$ (from (\ref{eq:tv3})). 
%%%%%%%%%%%%%%
\begin{table}[t!]
\caption{ITKM algorithm}
\begin{center}
\begin{tabular}{ll}
\hline\hline
 Given: $j$, $\widehat{\mathbf{s}}_\mathrm{g}^j$, $\mathbf{D}^0=\mathbf{D}^{j-1}\in\mathbb{R}^{n\times Q}$, $T$, and $h=1$ \\ \hline
  Repeat until convergence: \\
  1. Find dictionary indices per (\ref{eq:itkmSparsity4})\\
   \hspace{4ex} $K(\mathbf{D}^{h-1},\mathbf{y}_i)=\underset{|K|=T}\max\|{\mathbf{D}_K^{h-1}}^\mathrm{T}\mathbf{y}_{i}\|_1$, \\
  \hspace{4ex} with $\mathbf{y}_i=\mathbf{R}_i\widehat{\mathbf{s}}_\mathrm{g}^j-\mathbf{1}\overline{x}_i^j$. \\
  2. Update dictionary per (\ref{eq:itkmSparsity11}) using coefficient indices $K$\\
   \hspace{4ex}   $\mathbf{d}^{h}=\lambda_l\sum_{i:l\in K(\mathbf{D}^{h-1},\mathbf{y}_i)}\text{sign}\big({\mathbf{d}_l^{h-1}}^\mathrm{T}\mathbf{y}_{i}\big)\mathbf{y}_{i}$ \\
   \hspace{3ex} with $\lambda_l$ such that $\|\mathbf{d}_l^h\|_2=1$. \\
   \hspace{2ex} $h=h+1$ \\
   Ouput: $\mathbf{D}^j=\mathbf{D}^h$\\
\hline\hline
\end{tabular}
\label{algo:itkm}
\end{center}
\end{table}
%%%%%%%%%%%%%%%%%
%%%%%%%%%%%%%%%
%%%%%%%%%%%%%%%%
\par For the slowness maps, pixels are 1 km square and sampled by $M=2016$ straight-rays between 64 sensors, shown in Fig. \ref{fig:mapsSampling}(c). The travel times $\mathbf{t}$ is calculated by numerically integrating along these ray paths. The 2D valid-region for the slowness map estimates is determined for the LST by a dilation operation with a patch template along the outermost ray paths in Fig. \ref{fig:mapsSampling}(c). The conventional tomography valid region is bounded by the outermost pixels along the ray paths. The conventional tomography valid region is used for error calculations for both conventional and LST.

To avoid overfitting during dictionary learning (Table \ref{algo:itkm}), we exclude patches if more than 10 \% of the pixels are not sampled by ray paths. This heuristic works well and we have not investigated dictionary learning from incomplete information. The RMSE (ms/km) of the estimates is calculated by
%\begin{linenomath*}
\begin{equation}
\begin{aligned}
\text{RMSE}= \sqrt{\frac{1}{NP}\sum_\mathrm{n}^N\sum_p^P\big(s_{\mathrm{n}p}'-s_{\text{est.},\mathrm{n}p}'\big)^2},
\label{eq:rmse}
\end{aligned}
\end{equation}
%\end{linenomath*}
where $s_{\text{est.},\mathrm{n}p}'$ is $\widehat{\mathbf{s}}_\mathrm{s}'$, $\widehat{\mathbf{s}}_\mathrm{g}'$, or $\widehat{\mathbf{s}}_\mathrm{TV}'$ for the $\mathrm{n}$-th pixel location, and the $p$-th trial. For $\sigma_\epsilon=0$, $P=1$. The RMSE is printed on the 1D/2D slowness estimates in Figs.~5--12.

%%%%%%%%%%%%%%%%%%%%%%

%%%%%%%%%%%%%%%%%%%%%%%%
\subsection{Without travel time error}
We first simulate travel times without errors ($\sigma_\epsilon=0$) to obtain best-case results for the tomography methods. See Figs. \ref{fig:checker_noNoise} and \ref{fig:smoothDiscon_noNoise} for results from LST, conventional, and TV tomography. For the LST we invert the travel time using the algorithm in Table \ref{algo:patchSparse} both with and without dictionary learning. For the case with no dictionary learning, the dictionary $\mathbf{D}$ is either the overcomplete Haar wavelet dictionary or the DCT (see Fig. \ref{fig:dct}). RMSE performance is in Table \ref{table:rmse}. The performance of LST using prescribed dictionaries is related to previous works using prescribed dictionaries\cite{loris2007,loris2010,charlety2013,fang2014}.

\subsubsection{Inversion parameters for $\sigma_\epsilon=0$}
\label{sec:noiseFree}
The LST tuning parameters are $\lambda_1$, $\lambda_2$, $n$, $Q$, and $T$. The sensitivity of the LST solutions with dictionary learning to $\lambda_1$ and $\lambda_2$ are shown in Fig.~\ref{fig:noNoise_parmVary}(c,f) for nominal values of $n=100$, $Q=150$, $T=1$ for the checkerboard and $T=2$ for the smooth-discontinuous. With the prior $\sigma_\epsilon=0$\ s, we expect the best value of $\lambda_1=0$\ $\text{km}^2$ (from (\ref{eq:mapGlobal2})). From (\ref{eq:mapGlobal2}), (\ref{eq:mapLocal4}) $\sigma_\mathrm{g}^2$ is proportional to the variance of the true slowness. For the checkerboard (smooth-discontinuous) is $\sigma_\mathrm{g}=0.10$ ($0.05$)~$\text{s/km}$. We assume the slowness patches are well approximated by the sparse model (\ref{eq:mapLocal1}), and expect $\sigma_p^2\ll\sigma_\mathrm{g}^2$. Hence we expect the best value of $\lambda_2$ (from (\ref{eq:mapLocal4})) to be small. It is shown for both the checkerboard (Fig.~\ref{fig:noNoise_parmVary}(c)) and smooth-discontinuous maps (Fig.~\ref{fig:noNoise_parmVary}(f)) that the best RMSE for the LST with dictionary learning is obtained when $\lambda_1=0$\ $\text{km}^2$ and $\lambda_2=0$, though the LST exhibits low sensitivity to these values and recovers well the true slowness for a large range of values. We show the effect varying the sparsity $T$ on LST RMSE performance, relative to the true slowness per (\ref{eq:rmse}), for the nominal LST parameters in Fig.~\ref{fig:errorCurves}(b). The values used for $T$ were chosen based on the minimum error for these curves, though LST performance exceeds conventional and TV performance for a wide range of $T$.

For the LST with the Haar wavelet and DCT dictionaries (both $Q=169$, $n=64$, since Haar wavelet dimensions power of 2), the best performance by minimum RMSE was achieved with $\lambda_1=0$\ $\text{km}^2$, $\lambda_2=0$, and $T=5$ for the checkerboard and $T=2$ for smooth-discontinuous maps. 

For conventional tomography, there are several methods for estimating the best values of the regularization parameters $L$ and $\eta$, but the methods not always reliable \cite{bodin2009,aster2013}. To find the best parameters, we systematically varied the values of $L$ and $\eta$ (see Fig.~\ref{fig:noNoise_parmVary}(a,d)). The minimum RMSE for conventional tomography was obtained by $L=10$ km and $\eta=0.1\ \text{km}^2$ for both the checkerboard and smooth-discontinuous maps.

Similarly, for TV tomography, the values best values of the tuning parameters $\lambda_1$ and $\lambda_\mathrm{TV}$, were systematically varied the values of $\lambda_1$ and $\lambda_\mathrm{TV}$ (see Fig.~\ref{fig:noNoise_parmVary}(b,e)). The minimum RMSE for TV tomography was obtained by $\lambda_1=1~\text{km}^2$ and $\lambda_\mathrm{TV}=.01~\text{s}$ for both the checkerboard and smooth-discontinuous maps.

%%%%%%%%%%%%%%%%%
\begin{figure}
%\centering
\hspace{-0.3cm}\includegraphics[width=9.2cm]{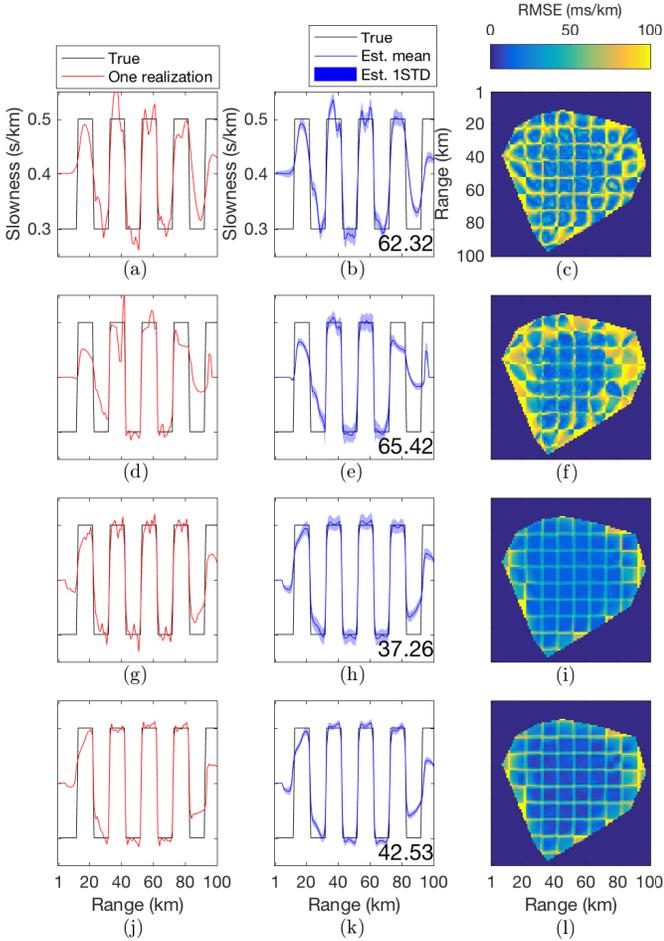}
\caption{Conventional, TV, and LST tomography results for checkerboard map (Fig. \ref{fig:mapsSampling}(a)) with 100 realizations of Gaussian travel time error (STD 2\% mean travel time): 1D slice of inversion for one noise realization against true slowness, 1D slice of mean from over all noise realizations with STD of estimate against true slowness $\mathbf{s}'$, and 2D RMSE of estimates over noise realizations. (a--c) conventional tomography $\widehat{\mathbf{s}}'_\mathrm{g}$, (d--f)  TV regularization $\widehat{\mathbf{s}}_\text{TV}'$, (g--i) LST $\widehat{\mathbf{s}}_\mathrm{s}'$ with dictionary learning with $\lambda_1=2 ~\text{km}^2$, and (j--l) $\lambda_1=7 ~\text{km}^2$. RMSE (ms/km), per (\ref{eq:rmse}), is printed on 1D errors.}
\label{fig:checker_noise}
\end{figure}
%%%%%%%%%%%%%%%%%%%%
\subsubsection{Results for $\sigma_\epsilon=0$}
\label{sec:noiseFree_results}
While the discontinuous shapes in the Haar dictionary are similar to the discontinuous content of the checkerboard image, the local features in the higher order Haar wavelets overfit the rays where the ray sampling density is poor (see Fig.\ \ref{fig:mapsSampling}(d)). The performance of the Haar wavelets is better for the smooth-discontinuous slowness map (Fig.~\ref{fig:smoothDiscon_noNoise}(g--i)) than for the checkerboard (Fig. \ref{fig:checker_noNoise}(e,f)). As shown in Fig.~\ref{fig:smoothDiscon_noNoise}(g--i), the Haar wavelets add false high frequency structure to the slowness reconstruction but the trends in the smooth-discontinuous features are well preserved. The inversion performance of the DCT transform (Fig. \ref{fig:checker_noNoise}(g,h) and Fig. \ref{fig:smoothDiscon_noNoise}(j--l)) is better than the Haar wavelets for both cases, but matches less closely the discontinuous slowness features, as the DCT atoms are smooth. The smoothness of the DCT atoms better preserve the smooth slowness structure. 

The LST with dictionary learning (Fig. \ref{fig:checker_noNoise}(i,j) and Fig.~\ref{fig:smoothDiscon_noNoise}(m--o)) achieves the best overall fit to the true slowness, recovering nearly exactly the true slownesses. As in the other cases, the performance degrades near the edges of the ray sampling, where the ray density is low, but high resolution is maintained across a large part of the sampling region. The RMSE of the LST with the Haar wavelet dictionary for the checkerboard (Table \ref{table:rmse}) is greater than for conventional tomography, although a better qualitative fit to the true slowness is observed with the LST. Both in the case of the checkerboard and smooth-discontinuous maps, the TV obtained the highest RMSE, though in regions of the map where the ray sampling was dense, the discontinuous and constant features were well recovered (Fig.~\ref{fig:checker_noNoise}(c,d) and Fig.~\ref{fig:smoothDiscon_noNoise}(d--f)), as expected for TV regularization.

The true variances $\sigma_\epsilon$, $\sigma_\mathrm{g}$, and $\sigma_p$ in the LST regularization parameters $\lambda_1$ and $\lambda_2$ provide best LST performance (see Fig.~\ref{fig:noNoise_parmVary}(c,f)), whereas the true variances for conventional tomography, $\sigma_\epsilon$ and $\sigma_\mathrm{c}$, do not correspond to the best solutions (see Fig.~\ref{fig:noNoise_parmVary}(a,d)). The noise-free prior, $\eta=0\ \text{km}^2$, gives erratic inversions, and the conventional tomography is better regularized by $\eta=0.1\ \text{km}^2$. The noise-free prior also gives erratic results from TV (see Fig.~\ref{fig:noNoise_parmVary}(b,e)).

A variety of checkerboard and smooth-discontinuous slowness maps with different geometries were generated to more fully test the tomography methods. The results of these tests are summarized in Table \ref{table:rmse} (and cases with travel time error are shown in Fig.~\ref{fig:mc}). Different checkerboard maps were generated by varying the size horizontal and vertical of the checkerboard boxes from 5 to 20 pixels (256 permutations), and different smooth-discontinuous maps were generated by varying the location of the left edge (from 32 to 62 pixels) and width (from 4 to 10 pixels) of the discontinuity (217 permutations). From each of these sets, 100 slowness maps were randomly chosen for simulation. Inversions without travel time errors were performed using conventional, TV, and LST (with dictionary learning) tomography using the nominal parameters from the aforementioned test cases, corresponding to the slowness maps in Fig.~\ref{fig:mapsSampling}(a,b). LST obtains lower RMSE than TV or conventional for all simulations for both varied checkerboard and smooth-discontinuous maps.
%%%%%%%%%%%%%%%%%%%%%%
\begin{figure*}
\centering
\includegraphics[width=12.3cm]{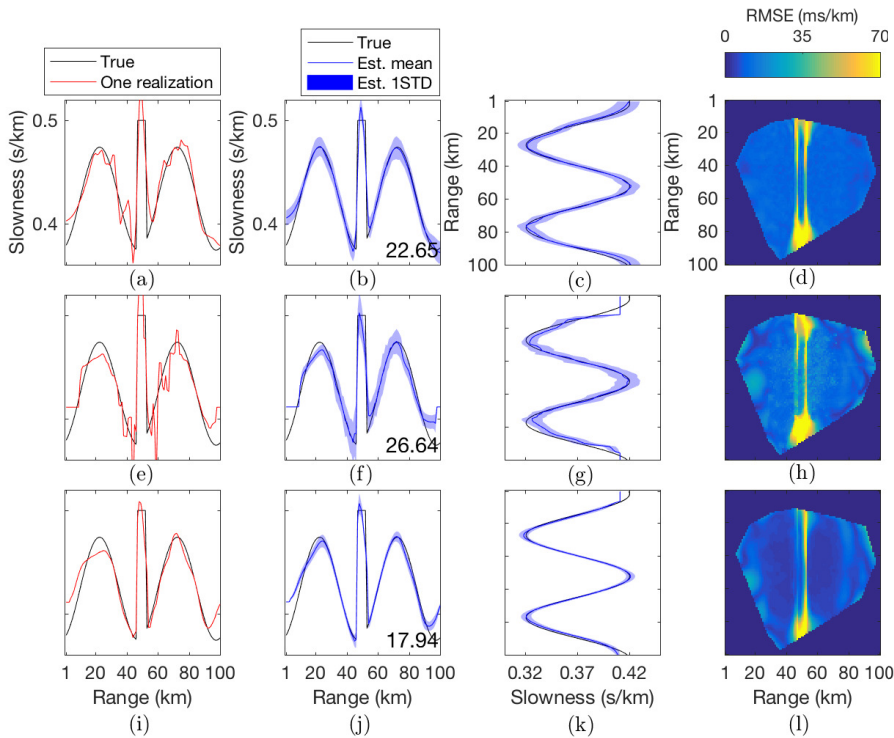}\caption{Conventional, TV, and LST tomography results for smooth-discontinuous map (Fig. \ref{fig:mapsSampling}(b)) with 100 realizations of Gaussian travel time error (STD 2\% mean travel time). Same format as Fig. 8, except vertical 1D slice of inversion is included, and only one LST case (i--l) $\widehat{\mathbf{s}}'_\mathrm{s}$ with dictionary learning. RMSE (ms/km), per (\ref{eq:rmse}), is printed on 1D errors.}
\label{fig:smoothDiscon_noise}
\end{figure*}
%%%%%%%%%%%%%%%%%%%%%%%
\subsection{With travel time error}
We also simulate the inversions with travel time errors. We consider the case when $\sigma_\epsilon=0.02\bar{t}$, or the uncertainty is 2\% of the mean travel time, which is similar to the model implemented in \cite{hawkins2015}. For each true slowness map and method, we run the inversions for $100$ realizations of noise $\mathcal{N}\big(0,\sigma_\epsilon\big)$ (also the random initialization of $\mathbf{D}$ also changes $100$ times for LST) and summarize the statistics of the results. The noise simulation results for conventional, TV, and LST tomography are in Figs. \ref{fig:checker_noise} and \ref{fig:smoothDiscon_noise}. The RMSE for both approaches, calculated by (\ref{eq:rmse}) with $P=100$, are in Table \ref{table:rmse}.

\subsubsection{Inversion parameters for $\sigma_\epsilon=0.02\bar{t}$}
The sensitivity of the LST solutions with dictionary learning to $\lambda_1$ and $\lambda_2$ are shown in Fig.~\ref{fig:noise_parmVary}(c,f) for nominal values of $n=100$, $Q=150$ and $T=2$ (per Fig.~\ref{fig:errorCurves}) for both the checkerboard and smooth-discontinuous maps. With the prior $\sigma_\epsilon=0.02\bar{t}$, we expect for the checkerboard map ($\sigma_\epsilon=0.27$\ s, $\sigma_\mathrm{g}=0.10\ \text{s/km}$) the best value of $\lambda_1\approx7.5$\ $\text{km}^2$, and for the the smooth-discontinuous ($\sigma_\epsilon=0.28$\ s, $\sigma_\mathrm{g}=0.05\ \text{s/km}$) the best value of $\lambda_1\approx28.3$\ $\text{km}^2$ (from (\ref{eq:mapGlobal2})). We use $\lambda_1=2\ \text{km}^2$ for the checkerboard (Fig.~\ref{fig:checker_noise}(g--i)) and $\lambda_1=10\ \text{km}^2$ for the smooth-discontinuous map (Fig.~\ref{fig:smoothDiscon_noise}(i--l)), and achieve lower RMSE than $\lambda_1=1\ \text{km}^2$ (see Fig.~\ref{fig:noise_parmVary}(c,f)). Although we expect the true values $\sigma_\mathrm{g}$ to decrease over the LST iterations, prior values of $\sigma_\mathrm{g}$ proportional to the variance of the true slowness work well. It is further shown in Fig.~\ref{fig:noise_parmVary} that, as in the the noise-free case (Sec.\ \ref{sec:noiseFree}), the LST recovers well the true slowness for a large range of values. 

For the LST with the Haar wavelet and DCT dictionaries, the best performance by minimum RMSE was achieved with $\lambda_1=5$\ $\text{km}^2$, $\lambda_2=0$, and $T=5$ for the checkerboard and $T=2$ for smooth-discontinuous maps. For conventional tomography, the best values by minimum RMSE were $L=6~\text{km}$ and $\eta=10\ \text{km}^2$ for the checkerboard and $L=12~\text{km}$ and $\eta=10\ \text{km}^2$ for the smooth-discontinuous slowness maps. For TV tomography, the best values by minimum RMSE were $\lambda_1=5~\text{km}^2$ and $\lambda_\mathrm{TV}=0.02~\text{s}$ for both the checkerboard and smooth-discontinuous slowness maps.

Considering again the influence of the choice of $\lambda_1$ and $\lambda_2$ on the LST with dictionary learning in Fig.~\ref{fig:noise_parmVary}(c,f), since $n=100$ the case $\lambda_2=100$ gives equal weight to the global $\widehat{s}_\mathrm{g,n}$ and patch slowness $b_\mathrm{n}s_\mathrm{p,n}$ in (\ref{eq:mapLocal8}). The LST obtains similar results best conventional estimates (see Fig.~\ref{fig:noise_parmVary}) for the checkerboard with $\lambda_1=1$ and $\lambda_2=100$ and for the smooth-discontinuous map with $\lambda_1=10$ and $\lambda_2=500$, though these parameters are suboptimal. Though in this case, the sparsity regularization of the patches has an effect similar to the conventional damping and smoothing regularization from (\ref{eq:classicMAP_solution1}), there is no direction relationship between the regularization in (\ref{eq:classicMAP_solution1}) and the sparsity and dictionary learning in (\ref{eq:mapLocal1}).
%%%%%%%%%%%%%
\begin{table*}[t]
\caption{Conventional, TV, and LST tomography RMSE performance. Bold entries are lowest error. }
\label{table:rmse}
\centering
\begin{tabular}{ l l | c|c|c|c|c|c|c}
\hline
& \multicolumn{8}{ c }{RMSE (ms/km)} \\ \cline{2-9}
& \multicolumn{4}{c|}{Checkerboard} & \multicolumn{4}{c}{Smooth-discon.} \\ \cline{2-9}
& \multicolumn{2}{c|}{Nominal (Fig.~\ref{fig:mapsSampling}(a))} & \multicolumn{2}{c|}{Varied} & \multicolumn{2}{c|}{Nominal (Fig.~\ref{fig:mapsSampling}(b))} & \multicolumn{2}{c}{Varied} \\ \cline{2-9}
\multicolumn{1}{c}{Error $\sigma_\epsilon=$} & $0^*$ & $0.02\bar{t}^{**}$ & $0$ & $0.02\bar{t}^{\ddagger}$ & $0^\dagger$ & $0.02\bar{t}^{\dagger\dagger}$ & $0$ & $0.02\bar{t}^{\ddagger}$ \\
\hline\hline
Conventional & 56.92 & 62.32 & 59.96 & 65.77 & 17.97 & 22.65 & 13.98 & 19.00\\
TV & 55.24 & 65.42 & 58.28 & 67.05 & 20.72 & 26.64 & 17.16 & 22.55\\
LST Haar & 60.97 & 68.96 & 59.71 & 66.51 & 15.66 & 23.07 & 14.00 & 18.68\\
LST DCT & 56.75 & 62.16 & 56.47 & 62.73 & 14.95 & 19.62 & 13.41 & 17.24\\
LST Adaptive & \bf{24.41} & \bf{37.26} & \bf{31.16} & \bf{37.14} & \bf{7.51} & \bf{17.94} & \bf{7.40} & \bf{14.62}\\ \hline
\multicolumn{8}{l }{\footnotesize{Estimated slownesses plotted in ${}^*$Fig.\ \ref{fig:checker_noNoise}, ${}^{**}$Fig.\ \ref{fig:checker_noise}, ${}^{\dagger}$Fig.\ \ref{fig:smoothDiscon_noNoise}, ${}^{\dagger\dagger}$Fig\ \ref{fig:smoothDiscon_noise}, and ${}^{\ddagger}$Fig\ \ref{fig:mc}.}}
\end{tabular}
\end{table*}
%%%%%%%%%%%%%%%%%%%%%%
\subsubsection{Results for $\sigma_\epsilon=0.02\bar{t}$}
The LST with dictionary learning (Fig. \ref{fig:checker_noise}(g--i) and Fig. \ref{fig:smoothDiscon_noise}(i--l)) achieves the best overall fit to the true slowness, relative to conventional tomography (Fig. \ref{fig:checker_noise}(a--c) and Fig. \ref{fig:smoothDiscon_noise}(a--d)) and TV tomography (Fig. \ref{fig:checker_noise}(d--f) and Fig. \ref{fig:smoothDiscon_noise}(e--h)) as evidenced by qualitative fit and RMSE (Table \ref{table:rmse}). In the case of the checkerboard, a higher value $\lambda_1=7~\text{km}^2$ is tested and shown in Fig. \ref{fig:checker_noise}(j--l). Because of the increased damping, the standard deviation (STD) of the estimate (Fig. \ref{fig:checker_noise}(k)) is less than the case $\lambda_1=2\ \text{km}^2$ (Fig. \ref{fig:checker_noise}(h)), and fit to true profile is improved. 

We also simulate inversion for a variety of checkerboard and smooth-discontinuous slowness maps with different geometries (per Sec.~\ref{sec:noiseFree_results}), with travel time error. The results of these tests are summarized in Table~\ref{table:rmse} and Fig.~\ref{fig:mc}. Inversions with 10 realizations of Gaussian travel time error ($\sigma_\epsilon=0.02\bar{t}$) were performed using conventional, TV, and LST (with dictionary learning) tomography using the nominal parameters from the aforementioned test cases, corresponding to the slowness maps in Fig.~\ref{fig:mapsSampling}(a,b). Fig.~\ref{fig:mc}(a,b) and Fig.~\ref{fig:mc}(c,d) show the results for the varied checkerboard and smooth-discontinuous maps. LST obtains lower RMSE than TV or conventional for all simulations for both varied checkerboard and smooth-discontinuous maps, as shown in Fig.~\ref{fig:mc}(a,c), a better subjective fit to the true slowness is also observed (Fig.~\ref{fig:mc}(b,d)).

\subsubsection{Convergence and run time}
\label{sec:convergence}
The algorithms were coded in Matlab. The LST algorithm (Table \ref{algo:patchSparse}) used 100 iterations for all cases and the ITKM (Table \ref{algo:itkm}) used 50 iterations. As an example of run time, the inversions with dictionary learning in Fig.~\ref{fig:checker_noNoise}(i,j) and Fig.~\ref{fig:smoothDiscon_noNoise}(m--o) took 5 min on a Macbook Pro 2.5 GHz Intel Core i7. The RMSE travel time error for the slowness model $\mathbf{s}_\mathrm{s}$ decreased over the iterations and converged within at most 50 iterations (see Fig. \ref{fig:errorCurves}). For the cases with travel time uncertainty, the RMSE approaches or falls only slightly below $\sigma_\epsilon$. Hence, the travel time data was not overfit.

\section{Conclusions}
We have derived a method for travel time tomography which incorporates a local sparse prior on patches of the slowness image, which we refer to as the LST algorithm. The LST can use predefined or learned dictionaries, though learned dictionaries gives improved performance. Relative to the conventional and TV tomography methods presented, the LST is less sensitive the regularization parameters. LST with sparse prior and dictionary learning can solve for both smooth and discontinuous slowness features in an image.

We considered the case of 2D surface wave tomography and, for the dense sampling configuration and synthetic images we used, obtained good recovery of true slowness maps with exclusively discontinuous features as well as smooth and discontinuous features using LST. The LST is relevant to other tomography scenarios where slowness structure is irregularly sampled, for instance in ocean \cite{verlinden15} and terrestrial \cite{annibale2013,rabenstein2017} acoustics.

%%%%%%%%%%%%%%%%%%%%%%%
\appendix
\label{sec:appendix}
\subsection{ITKM algorithm details}
\label{sec:appendixITKM}
The ITKM dictionary learning algorithm \cite{schnass2015} (see Table \ref{algo:itkm}) is derived from a `signed' K-means objective. In signed K-means, $T$-sparse coefficients $\mathbf{C}=[\mathbf{c}_1,...,\mathbf{c}_I]\in\mathbb{R}^{Q\times I}$ with $\mathbf{c}_i^T\in\{-1,1\}$ are assigned to training examples $\{\mathbf{y}_i,...,\mathbf{y}_I\}\in\mathbb{R}^{n\times I}$. The training examples in this paper obtained from patch $i$ as $\mathbf{y}_i=\mathbf{R}_i\mathbf{u}$, and centered. The minimization problem is
%\begin{linenomath*}
\begin{equation}
\begin{aligned}
\big\{\mathbf{C},\mathbf{D}\big\}=& \underset{\mathbf{D}}{\arg\min} \ \sum_i\underset{T, \mathbf{c}_i^T=\pm1}{\arg\min} \|\mathbf{y}_i-\mathbf{D}\mathbf{c}_{i}\|_2^2 \\
=&\underset{\mathbf{D}}{\arg\min}\sum_i\underset{T,\ c_i^t=\pm1}{\arg\min} \ \|\mathbf{y}_i-\sum_t\mathbf{d}_tc_{i}^t\|_2^2,
\end{aligned}
\label{eq:itkmSparsity1}
\end{equation}
%\end{linenomath*}
where $c_{i}^t$ is a non-zero coefficient and $\mathbf{d}_t$ is the corresponding dictionary atom. Expanding (\ref{eq:itkmSparsity1}) and requiring $\|\mathbf{d}_t\|_2^2=1$,
\begin{gather}
\begin{aligned}
\underset{\mathbf{D}}{\min}&\sum_i\underset{T,\ c_i^t=\pm1}{\min} \ \bigg\{ \|\mathbf{y}_i\|_2^2-2\sum_tc_i^t\mathbf{d}_t^\mathrm{T}\mathbf{y}_{i} +B\bigg\} \\
&= \|Y\|_\mathcal{F}^2 +B -2 \ \underset{\mathbf{D}}{\max}\sum_i\underset{|K|=T}{\max}\sum_t\mathrm{abs}\big(\mathbf{d}_t^\mathrm{T}\mathbf{y}_{i}\big) \\
&= \|Y\|_\mathcal{F}^2 +B -2 \ \underset{\mathbf{D}}{\max}\sum_i\underset{|K|=T}{\max}\|\mathbf{D}_K^\mathrm{T}\mathbf{y}_{i}\|_1,
\end{aligned} \raisetag{2.3\baselineskip}
\label{eq:itkmSparsity3}
\end{gather}
%\end{linenomath*}
where $B$ is a constant, and $K$ is the set of $T$ dictionary indices having the largest absolute inner product $\mathrm{abs}(\mathbf{d}_t\mathbf{y}_i)$, by
%\begin{linenomath*}
\begin{equation}
K(\mathbf{D},\mathbf{y}_i)=\underset{|K|=T}\max\|\mathbf{D}_K^\mathrm{T}\mathbf{y}_{i}\|_1.
\label{eq:itkmSparsity4}
\end{equation}
%\end{linenomath*}
In contrast to OMP\cite{pati93}, (\ref{eq:itkmSparsity4}) is the thresholding \cite{elad2010} solution to (\ref{eq:mapLocal1}). From (\ref{eq:itkmSparsity3}), the dictionary learning objective is 
%\begin{linenomath*}
\begin{equation}
\underset{\mathbf{D}}{\max}\sum_i\underset{|K|=T}{\max}\|\mathbf{D}_K^\mathrm{T}\mathbf{y}_{i}\|_1,
\label{eq:itkmSparsity5}
\end{equation}
%\end{linenomath*}
which finds $\mathbf{D}$ that maximizes the absolute norm of the $T$-largest responses from $K$. 
%%%%%%%%%%%%%%%%%%%%
\begin{figure*}[!t]
\centering
\subfloat{\includegraphics[width=5cm]{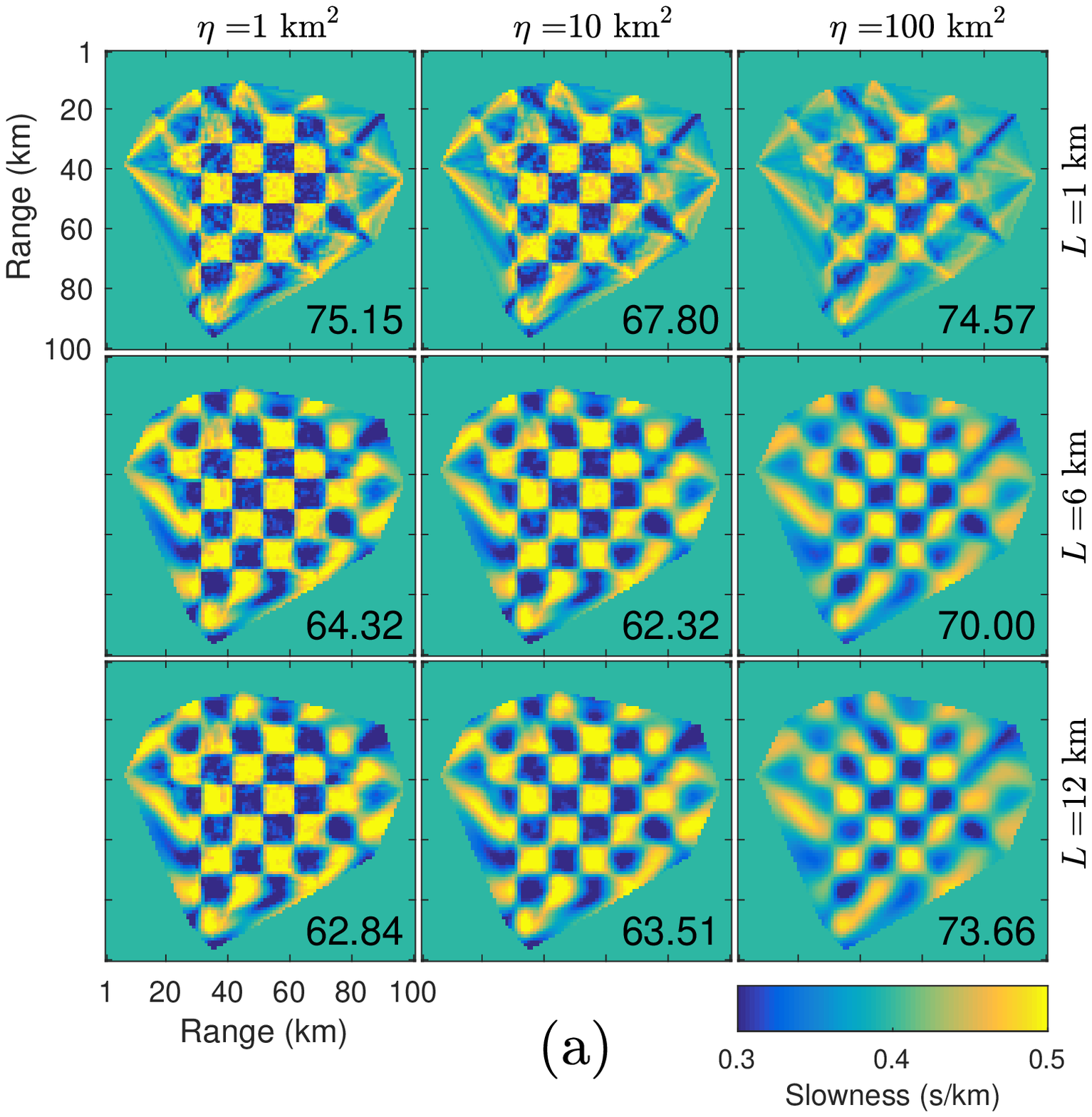}
\label{fig_first_case}}
\subfloat{\includegraphics[width=5cm]{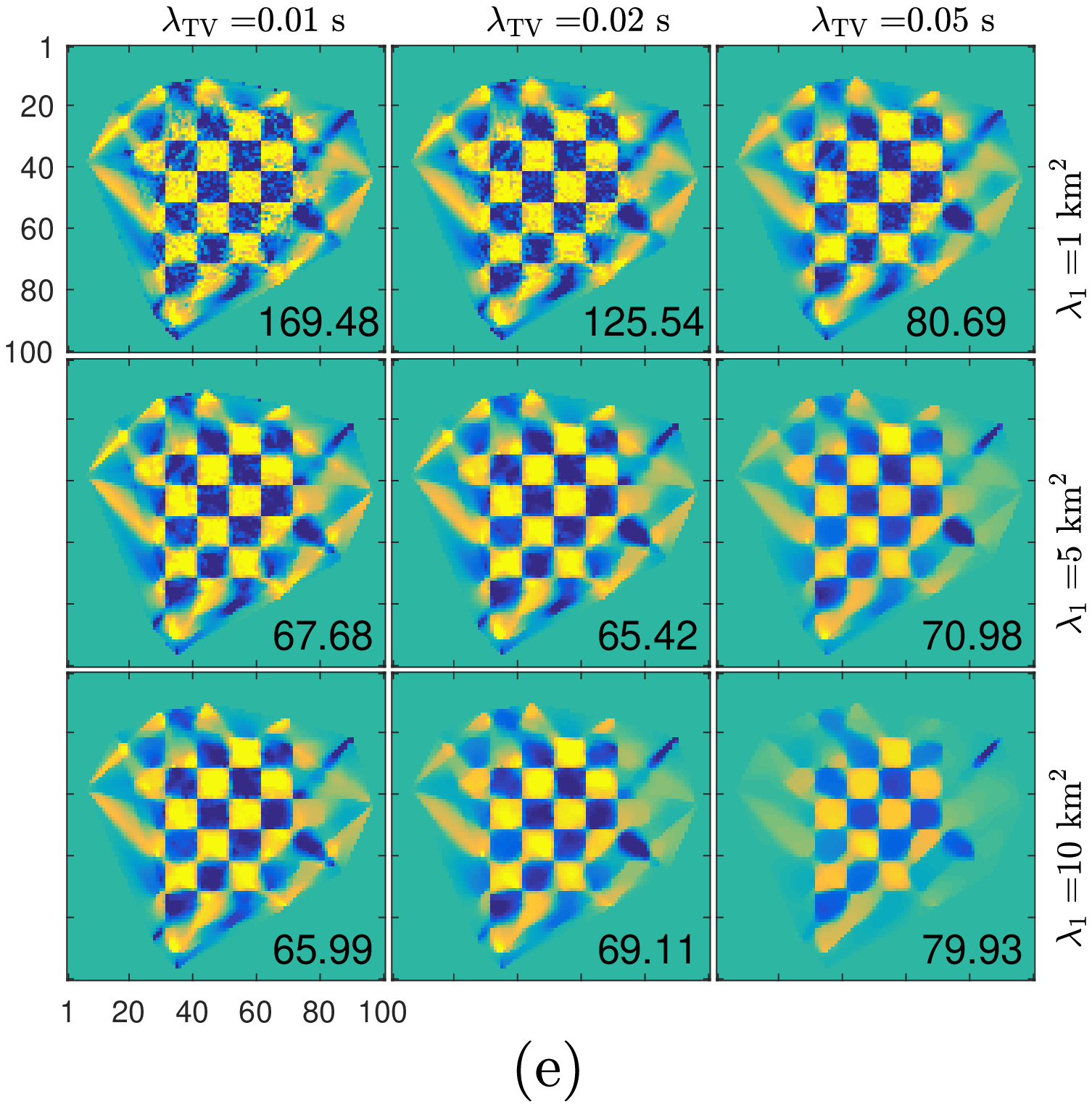}
\label{fig_second_case}}
\subfloat{\includegraphics[width=7.8cm]{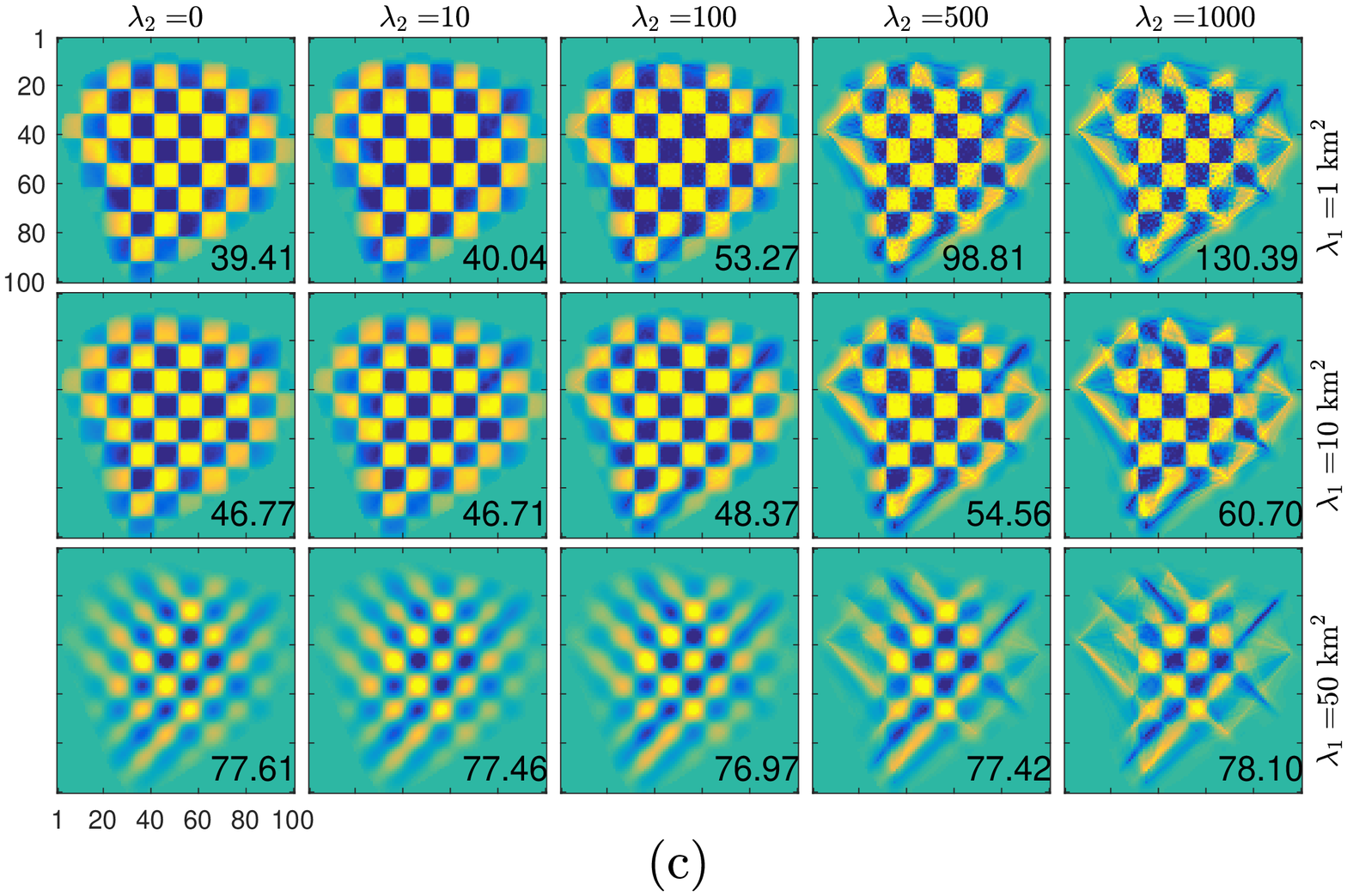}
\label{fig_second_case}}\vspace{-0.6cm}
\subfloat{\includegraphics[width=5cm]{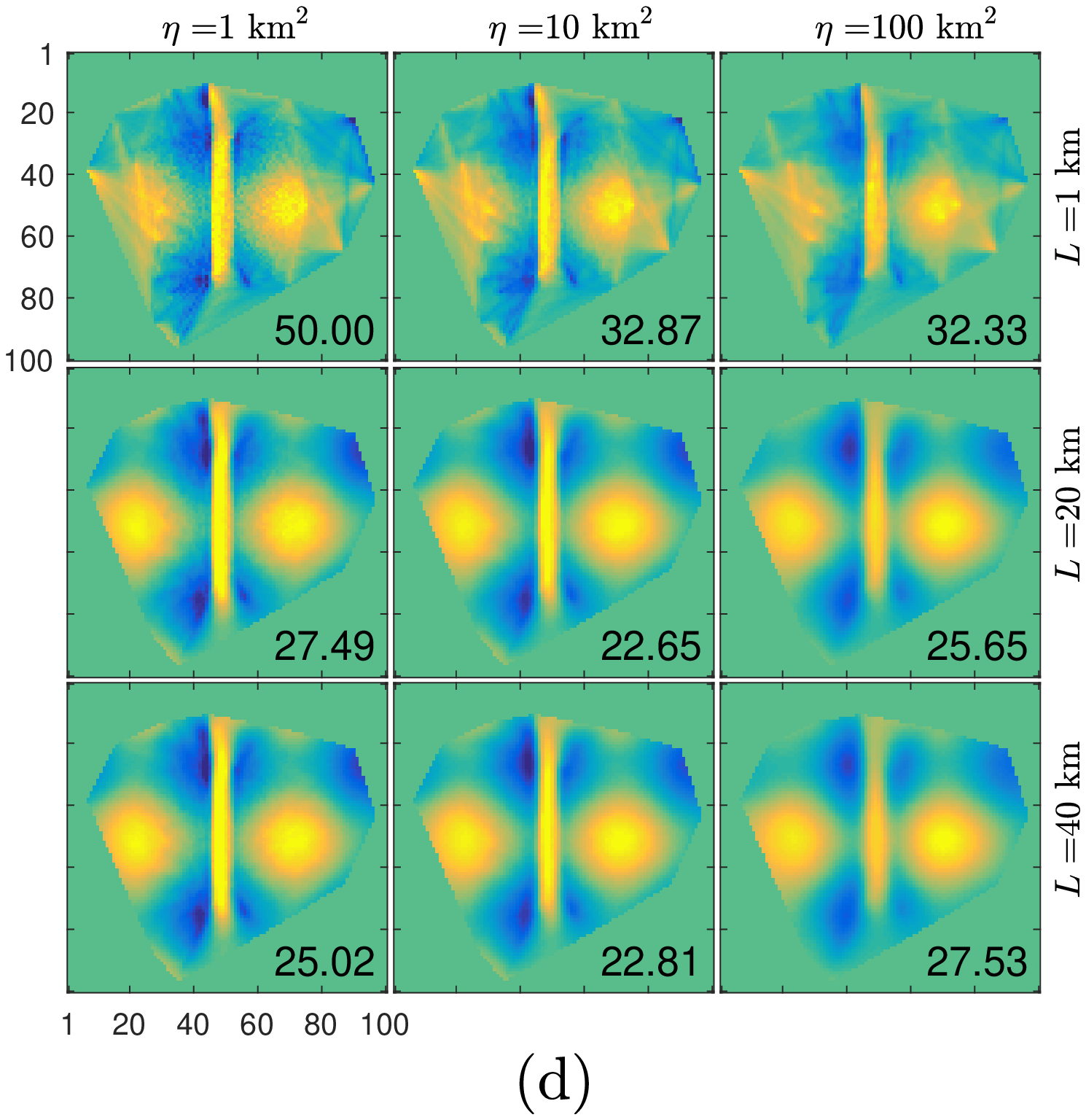}
\label{fig_first_case}}
\subfloat{\includegraphics[width=5cm]{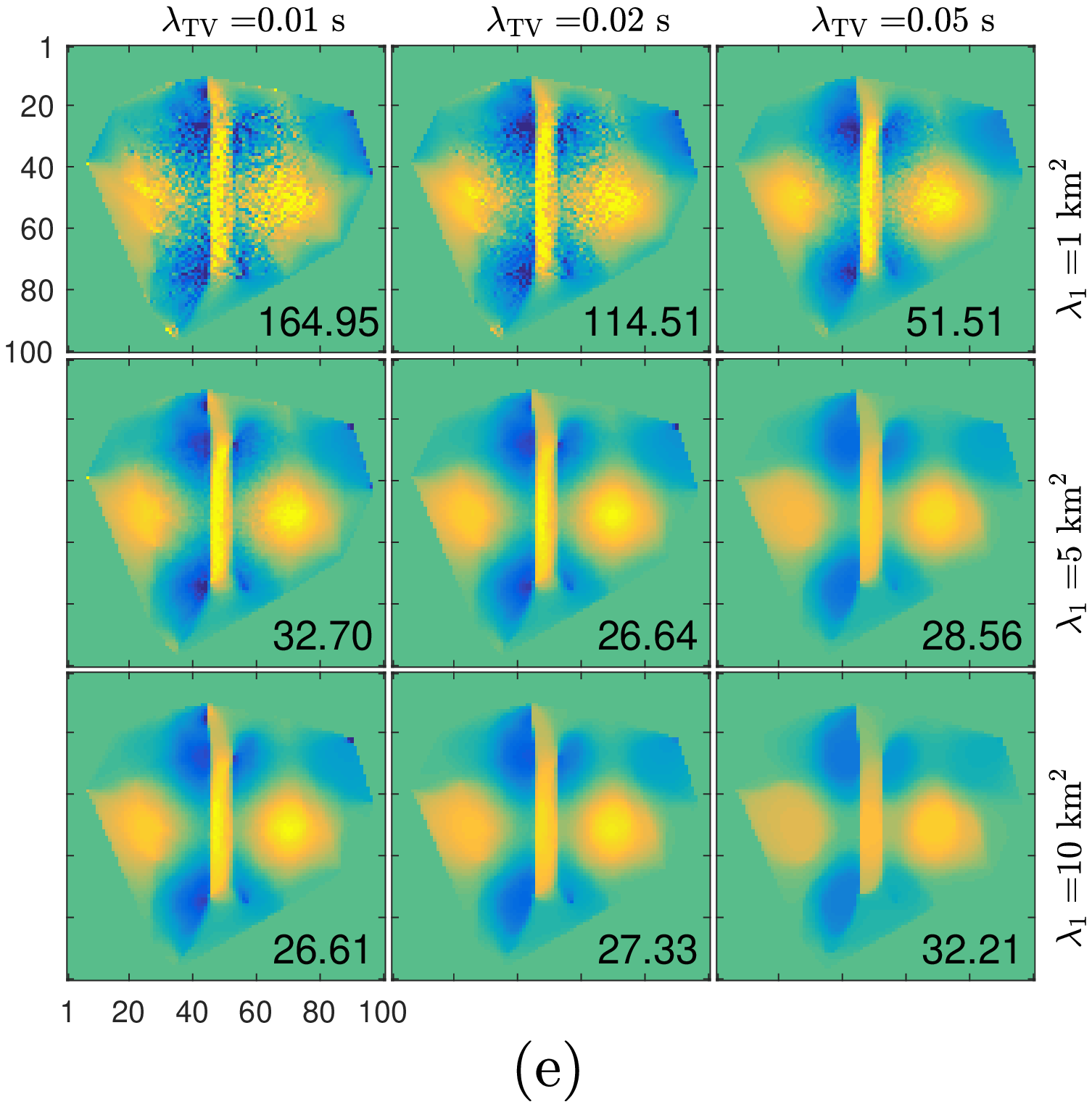}
\label{fig_second_case}}
\subfloat{\includegraphics[width=7.8cm]{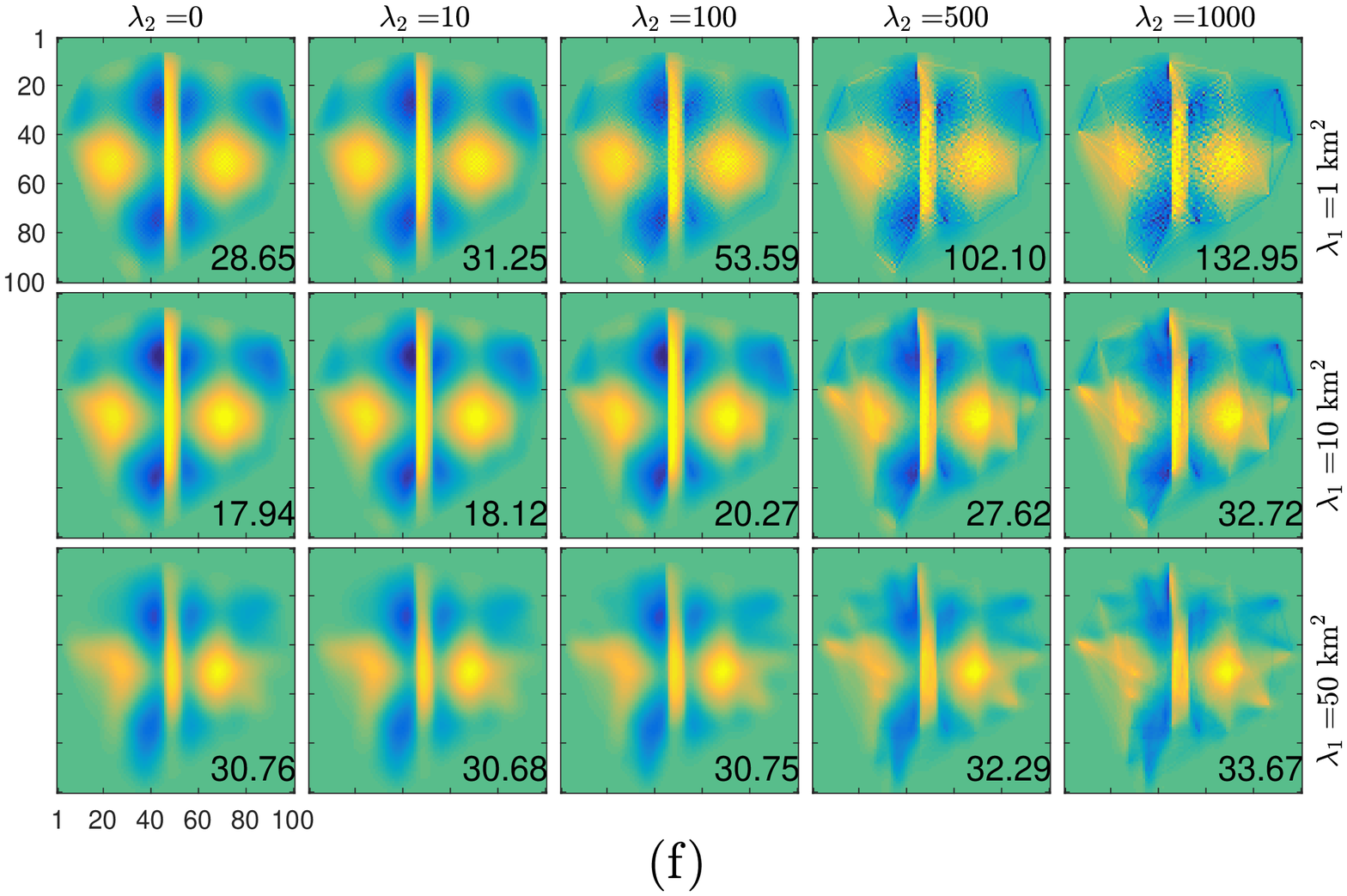}
\label{fig_second_case}}\\
\caption{Conventional, TV, and LST tomography results for different values of regularization parameters for (a--c) checkerboard (Fig. \ref{fig:mapsSampling}(a)) and (d--f) smooth-discontinuous (Fig. \ref{fig:mapsSampling}(b)) maps, for 100 realizations of Gaussian travel time error (STD 2\% mean travel time). (a,d) Conventional mean $\widehat{\mathbf{s}}'_\mathrm{g}$, effect of $L$ and $\eta$. (b,e) TV regularization mean $\widehat{\mathbf{s}}'_\text{TV}$, effect of $\lambda_1$ and $\lambda_\text{TV}$. (c,f) LST mean $\widehat{\mathbf{s}}'_\mathrm{s}$, effect of $\lambda_1$ and $\lambda_2$. RMSE (ms/km), per (\ref{eq:rmse}), is printed on 2D slownesses.}
\label{fig:noise_parmVary}
\end{figure*}
%%%%%%%%%%%%%%%%%%%%%%%
The ITKM obtains dictionary learning by solving (\ref{eq:itkmSparsity5}) as a two-step algorithm. First, $K$ is obtained from (\ref{eq:itkmSparsity4}). Then, the dictionary atoms are updated per
%\begin{linenomath*}
\begin{equation}
\underset{\mathbf{D}}{\max}\hspace{-2ex}\sum_{i:l\in K(\mathbf{D},\mathbf{y}_i)}\hspace{-3ex}\mathrm{abs}\big(\mathbf{d}_l^\mathrm{T}\mathbf{y}_{i}\big).
\label{eq:itkmSparsity6}
\end{equation}
%\end{linenomath*}
This optimization problem is solved using Lagrange multipliers, with the constraint $\|\mathbf{d}_l\|_2=1$. The Lagrangian function is
%\begin{linenomath*}
\begin{equation}
\Phi(\mathbf{d}_l,\lambda)=\sum_{i:l\in K(\mathbf{D},\mathbf{y}_i)}\hspace{-3ex}\mathrm{abs}\big(\mathbf{d}_l^\mathrm{T}\mathbf{y}_{i}\big)-\lambda\big(\mathbf{d}_l^\mathrm{T}\mathbf{d}_l-1\big),
\label{eq:itkmSparsity7}
\end{equation}
%\end{linenomath*}
with $\lambda$ the Lagrange multiplier. Differentiating (\ref{eq:itkmSparsity7}) gives
\begin{equation}
\frac{d\Phi}{d\mathbf{d}_l}=\hspace{-2ex}\sum_{i:l\in K(\mathbf{D},\mathbf{y}_i)}\hspace{-3ex}\mathrm{sign}\big(\mathbf{d}_l^\mathrm{T}\mathbf{y}_{i}\big)\mathbf{y}_{i}-2\lambda\mathbf{d}_l.
\label{eq:itkmSparsity8}
\end{equation}
%\end{linenomath*}
The stationary point of (\ref{eq:itkmSparsity7}), per (\ref{eq:itkmSparsity8}), gives the update for $\mathbf{d}_l$ as
%\begin{linenomath*}
\begin{equation}
\mathbf{d}_l^{new}=\lambda_l\hspace{-3ex}\sum_{i:l\in K(\mathbf{D}^{old},\mathbf{y}_n)}\hspace{-3ex}\text{sign}\big({\mathbf{d}_l^{old}}^\mathrm{T}\mathbf{y}_{i}\big)\mathbf{y}_{i},
\label{eq:itkmSparsity11}
\end{equation}
%\end{linenomath*}
where $\lambda_l=1/(2\lambda)$. The complexity of each ITKM iteration is dominated by matrix multiplication, of order $O(nQI)$, which is much less than K-SVD \cite{aharon2006} which for each iteration calculates the SVD of a $n\times I$ matrix $Q$ times.
%%%%%%%%%%%%%%%%%%%%%%
\begin{figure}[t!]
\centering
\subfloat{\includegraphics[width=4.5cm]{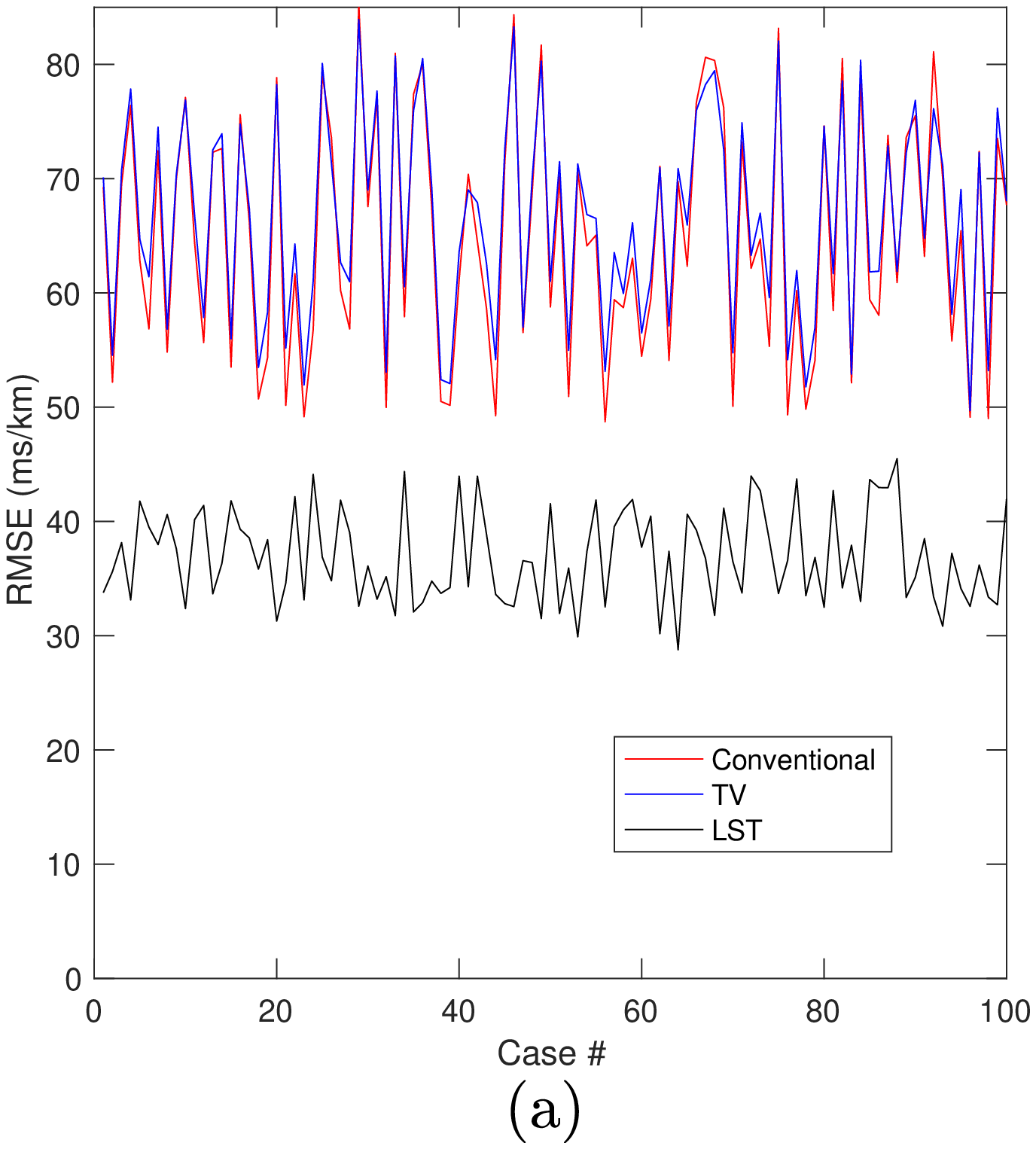}\hspace{-0.4cm}
\label{fig_first_case}}
\subfloat{\includegraphics[width=4.5cm]{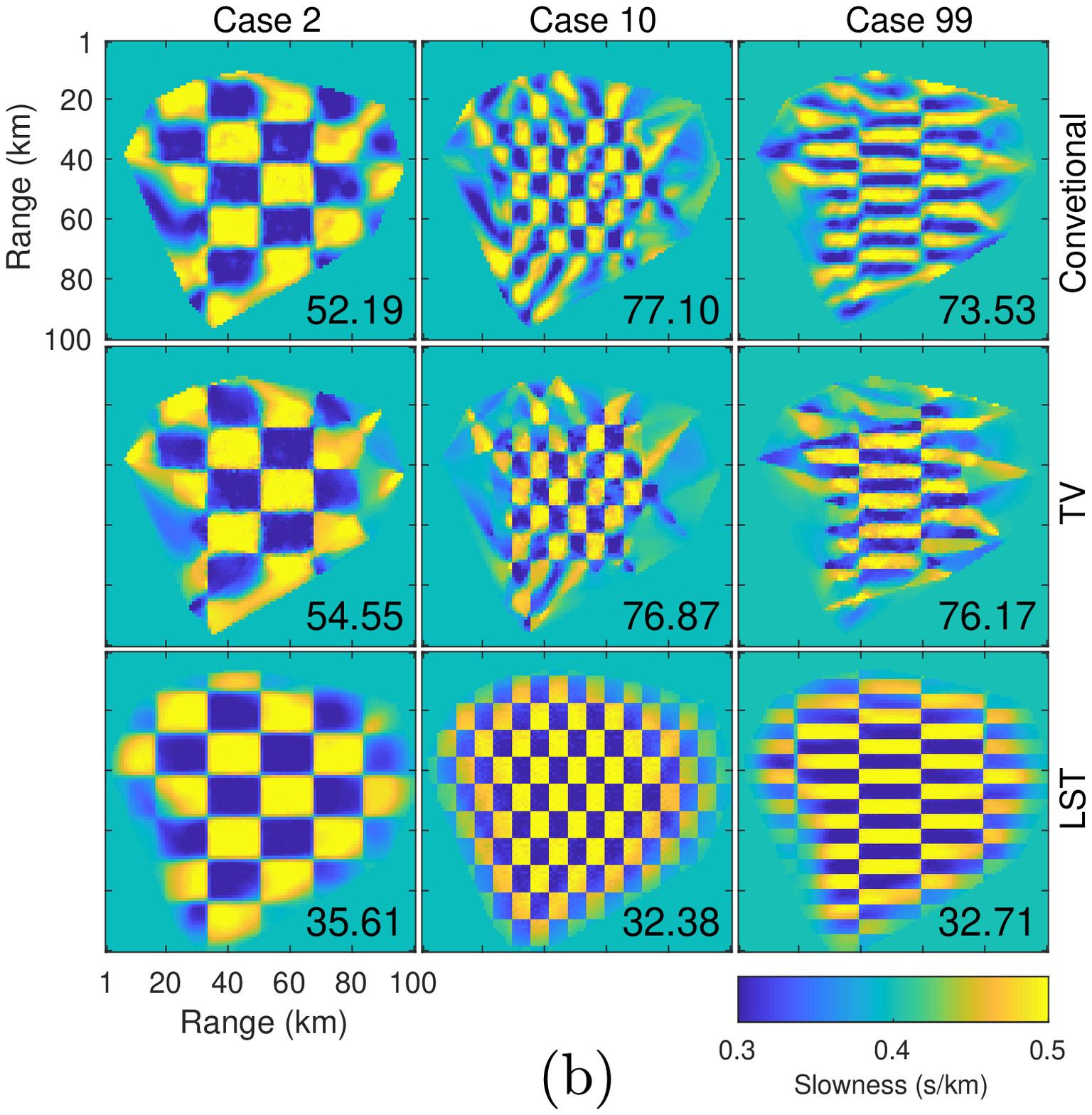}
\label{fig_second_case}}\vspace{-0.3cm}
\subfloat{\includegraphics[width=4.5cm]{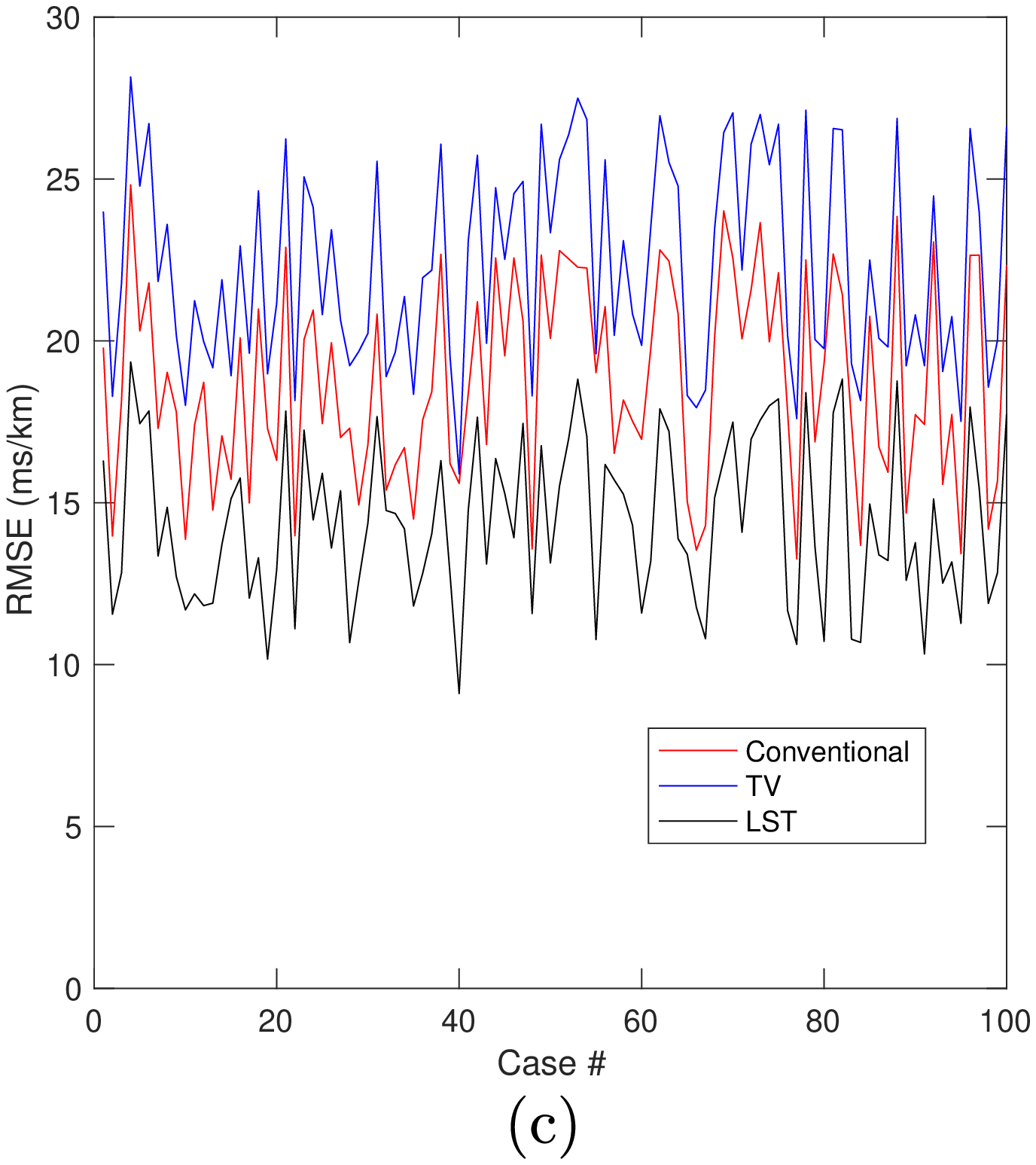}\hspace{-0.4cm}
\label{fig_first_case}}
\subfloat{\includegraphics[width=4.5cm]{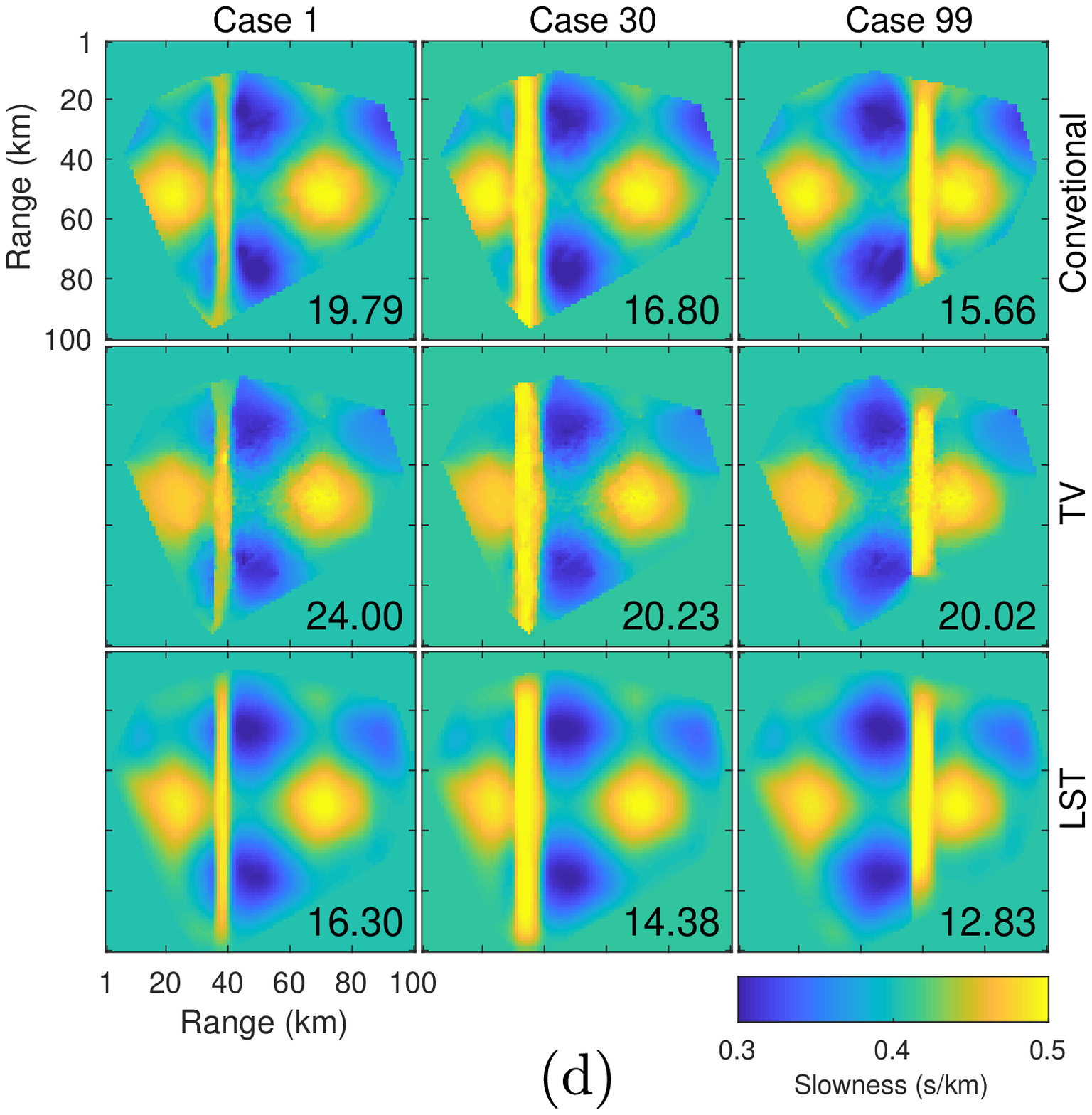}
\label{fig_second_case}}
\caption{Conventional, TV, and LST tomography results for 100 different checkerboard and smooth-discontinuous slowness map configurations and 10 realizations of travel time error ($\sigma_\epsilon=0.02\bar{t}$). (a,c) Comparison of RMSE error for 100 cases, and (b,d) mean slowness for 3 example cases from (a,c). RMSE (ms/km), per (\ref{eq:rmse}), is printed on 2D slownesses.}
\label{fig:mc}
\end{figure}
%%%%%%%%%%%%%%%%%%%%%%
\begin{figure}[t]
\hspace{-.3cm}
\subfloat[]{\includegraphics[width=4.6cm]{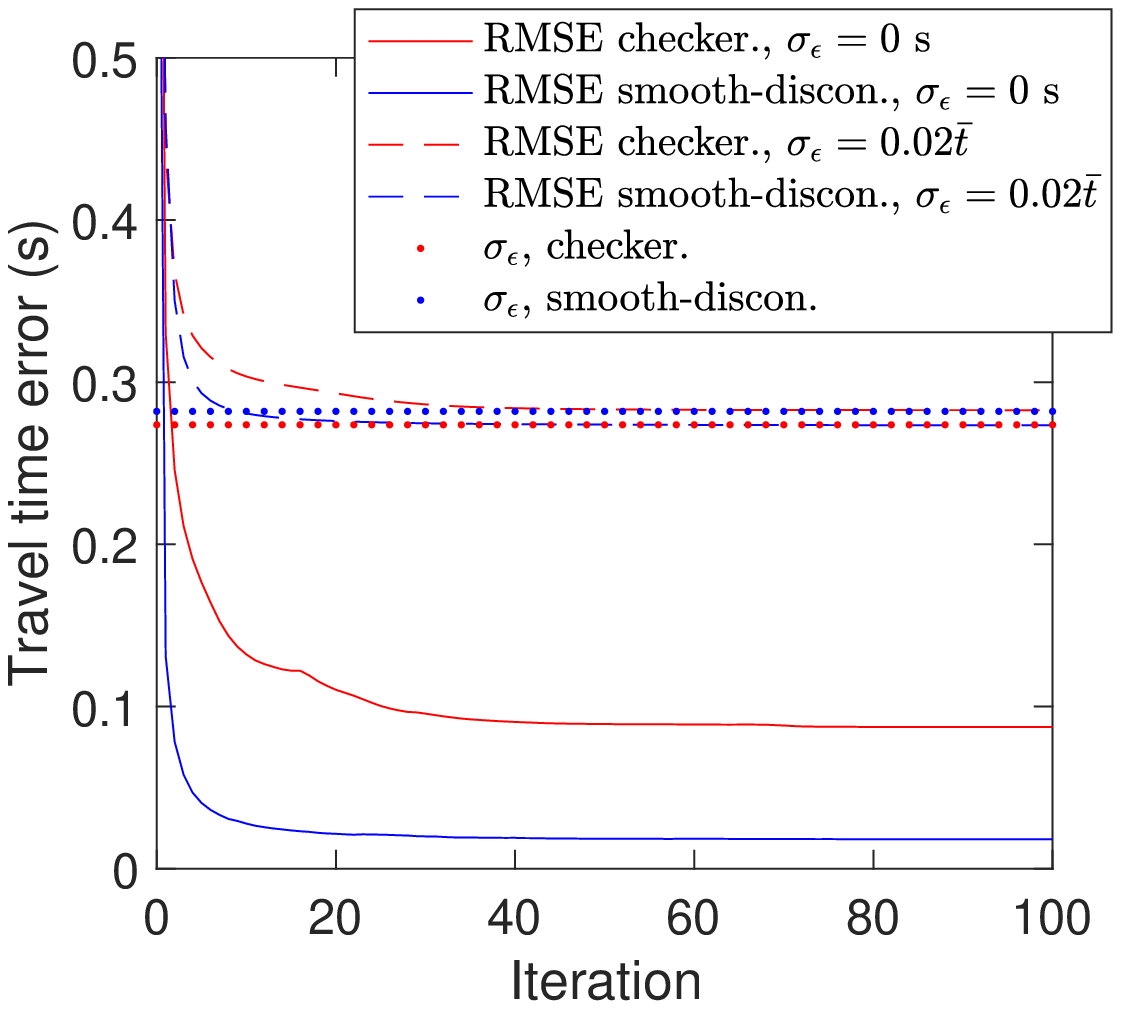}
\label{fig_first_case}}
\subfloat[]{\includegraphics[width=4.6cm]{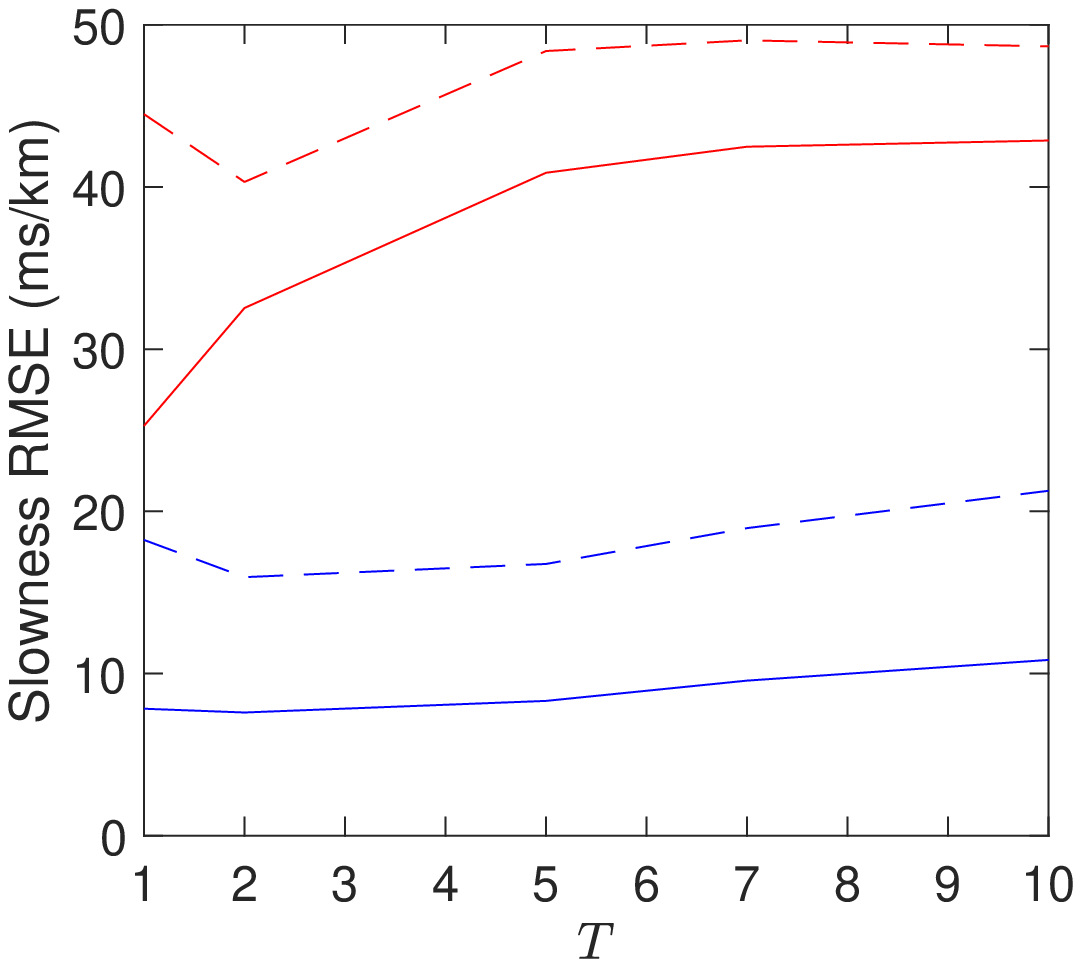}
\label{fig_second_case}}
\caption{(a) LST algorithm travel time RMSE convergence vs. iteration (Table \ref{algo:patchSparse}) and (b) slowness RMSE vs. sparsity $T$ with and without travel time error, with dictionary learning. Results shown with and without travel time error, corresponding to the checkerboard (Fig.~\ref{fig:checker_noNoise}(i,j), \ref{fig:checker_noise}(g--i)) and smooth-discontinuous (Fig.~\ref{fig:smoothDiscon_noNoise}(m--o), \ref{fig:smoothDiscon_noise}(i--l)) slowness maps.}
\label{fig:errorCurves}
\end{figure}

\ifCLASSOPTIONcaptionsoff
  \newpage
\fi

% trigger a \newpage just before the given reference
% number - used to balance the columns on the last page
% adjust value as needed - may need to be readjusted if
% the document is modified later
%\IEEEtriggeratref{8}
% The "triggered" command can be changed if desired:
%\IEEEtriggercmd{\enlargethispage{-5in}}

% references section

% can use a bibliography generated by BibTeX as a .bbl file
% BibTeX documentation can be easily obtained at:
% http://mirror.ctan.org/biblio/bibtex/contrib/doc/
% The IEEEtran BibTeX style support page is at:
% http://www.michaelshell.org/tex/ieeetran/bibtex/
%\bibliographystyle{IEEEtran}
% argument is your BibTeX string definitions and bibliography database(s)
%\bibliography{IEEEabrv,../bib/paper}
%
% <OR> manually copy in the resultant .bbl file
% set second argument of \begin to the number of references
% (used to reserve space for the reference number labels box)
\bibliographystyle{IEEEtran}
\bibliography{paperRefs_042618}

%\begin{thebibliography}{1}
%
%\bibitem{IEEEhowto:kopka}
%H.~Kopka and P.~W. Daly, \emph{A Guide to \LaTeX}, 3rd~ed.\hskip 1em plus
%  0.5em minus 0.4em\relax Harlow, England: Addison-Wesley, 1999.
%
%\end{thebibliography}

% biography section
% 
% If you have an EPS/PDF photo (graphicx package needed) extra braces are
% needed around the contents of the optional argument to biography to prevent
% the LaTeX parser from getting confused when it sees the complicated
% \includegraphics command within an optional argument. (You could create
% your own custom macro containing the \includegraphics command to make things
% simpler here.)
%\begin{IEEEbiography}[{\includegraphics[width=1in,height=1.25in,clip,keepaspectratio]{mshell}}]{Michael Shell}
% or if you just want to reserve a space for a photo:
%%%
%%%\begin{IEEEbiography}{Michael Bianco}
%%%Biography text here.
%%%\end{IEEEbiography}
%%%
%%%% if you will not have a photo at all:
%%%\begin{IEEEbiographynophoto}{Peter Gerstoft}
%%%Biography text here.
%%%\end{IEEEbiographynophoto}

% insert where needed to balance the two columns on the last page with
% biographies
%\newpage

%\begin{IEEEbiographynophoto}{Jane Doe}
%Biography text here.
%\end{IEEEbiographynophoto}

% You can push biographies down or up by placing
% a \vfill before or after them. The appropriate
% use of \vfill depends on what kind of text is
% on the last page and whether or not the columns
% are being equalized.

%\vfill

% Can be used to pull up biographies so that the bottom of the last one
% is flush with the other column.
%\enlargethispage{-5in}

% that's all folks
\end{document}